\documentclass[
 reprint,
 amsmath,amssymb,
 pra,superscriptaddress, nofootinbib
]{revtex4-2}
\usepackage[utf8]{inputenc}
\usepackage{graphicx}
\usepackage{bm}
\usepackage{stmaryrd}
\usepackage{dsfont}
\usepackage{tikz}
\usepackage{tkz-euclide}
\usepackage{amsthm}
\usepackage{mathtools}
\usepackage{hyperref}
\usepackage{braket}
\usepackage{pgfplots}
\usepackage{enumerate}
\usepackage[caption=false]{subfig}
\usepackage{booktabs}
\usepackage{xcolor}
\usepgfplotslibrary{external}
\DeclareMathOperator{\sgn}{sgn}

\begin{document}

\preprint{APS/123-QED}

\title{Ising on the donut:~Regimes of topological quantum error correction from statistical~mechanics}

\author{Lucas~H.~English}
 \email{lucas.english@sydney.edu.au}
\affiliation{%
Centre for Engineered Quantum Systems, School of Physics, University of Sydney, Sydney, NSW 2006, Australia
}%
\author{Sam Roberts}
\affiliation{
PsiQuantum, Brisbane, Australia
}
\author{Stephen~D.~Bartlett}
\affiliation{%
Centre for Engineered Quantum Systems, School of Physics, University of Sydney, Sydney, NSW 2006, Australia
}%
\author{Andrew~C.~Doherty}
\affiliation{%
Centre for Engineered Quantum Systems, School of Physics, University of Sydney, Sydney, NSW 2006, Australia
}%
\affiliation{
PsiQuantum, Brisbane, Australia
}
\author{Dominic~J.~Williamson}
\affiliation{%
Centre for Engineered Quantum Systems, School of Physics, University of Sydney, Sydney, NSW 2006, Australia
}%
\affiliation{
PsiQuantum, Brisbane, Australia
}%

\date{\today}

\begin{abstract}
Utility-scale quantum computers require quantum error correcting codes with large numbers of physical qubits to achieve sufficiently low logical error rates.  The performance of quantum error correction (QEC) is generally predicted through large-scale numerical simulations, used to estimate thresholds, finite-size scaling, and exponential suppression of logical errors below threshold. The connection of QEC to models from statistical mechanics provides an alternative tool for analysing QEC performance. However, predicting the behaviour of these models also requires large-scale numerical simulations, as analytic solutions are not generally known.  Here we exploit an exact mapping, from a toric code under bit-flip noise that is post-selected on being syndrome free to the exactly-solvable two-dimensional Ising model on a torus, to derive an analytic solution for the logical failure rate across its full domain of physical error rates.  In particular, this mapping provides closed-form expressions for the logical failure rate in four distinct regimes: the path-counting, below-threshold (ordered), near-threshold (critical), and above-threshold (disordered) regimes. Our framework places a number of familiar and long-standing numerical observations on firm theoretical ground.  It also motivates explicit ans\"{a}tze for the conventional QEC setting of non-post-selected codes whose statistical mechanics mappings involve random-bond disorder. Specifically, we introduce an effective tension model for the below-threshold regime, and a new scaling ansatz for the near-threshold regime, derived from an analysis of the domain wall energy cost distributions. By bridging statistical mechanics theory and quantum error correction practice, our results offer a new toolkit for designing, benchmarking, and understanding topological codes beyond current computational limits.
\end{abstract}

\maketitle

\section{Introduction}\label{sec:intro}

As quantum devices scale toward fault-tolerant quantum computing (FTQC), with the ambition of deploying millions of physical qubits to perform useful quantum computations with noisy components~\cite{10.1145/3723153}, quantum error correction (QEC) research has come to rely almost exclusively on large-scale numerical simulations to benchmark code performance. From early Monte Carlo studies of concatenated and surface codes~\cite{PhysRevA.74.052333,10.1063/1.1499754}, to more recent tensor network and Monte Carlo estimates of threshold values~\cite{PhysRevA.90.032326,PhysRevLett.103.090501,chubb2021generaltensornetworkdecoding} and finite-size scaling~\cite{WangChenyang2003Ctia, Watson_2014, Chubb_2021}, these numerical approaches have delivered indispensable insights into logical error suppression below-threshold and resource requirements for overhead optimization. Yet, there are few analytic treatments, the most notable being rigorous lower bounds on thresholds for concatenated codes \cite{Aharonov_1997, Preskill_1998, Knill_1998, Aliferis_2006} and upper-bound estimates via statistical mechanics constructions \cite{WangChenyang2003Ctia, PhysRevX.2.021004, PhysRevLett.103.090501, Xiao2024exactresultsfinite}. These treatments typically yield conservative bounds rather than precise performance predictions. As the system complexity of quantum computers grows beyond what even the most powerful supercomputers can faithfully simulate, analytic methods capable of delivering accurate, regime-spanning performance formulas will be essential to guide code design, decoder development, and resource allocation in the FTQC regime.

There is a rich connection between topological codes and classical statistical mechanics \cite{10.1063/1.1499754,WangChenyang2003Ctia,PhysRevX.2.021004,Watson_2014,Chubb_2021,PhysRevB.103.104306,Williamson2021On,Song2021Optimal,Zhu2023Nishimori,Xiao2024exactresultsfinite,PRXQuantum.6.010327}. Notably, explicit mappings have been established between logical failure rates of well-known topological codes, such as surface codes, and partition functions of well-known disordered statistical mechanical models \cite{10.1063/1.1499754,PhysRevX.2.021004}. A celebrated example is the correspondence between the surface code under a bit-flip error channel and the random-bond Ising model (RBIM) in two dimensions, featuring $\pm J$ couplings \cite{10.1063/1.1499754}. The order-disorder phase transition of the statistical mechanical model corresponds to the logical threshold of the code \cite{10.1063/1.1499754,Chubb_2021}. In the thermodynamic limit, for a physical error rate below this critical value, decoding success occurs with certainty. Above this critical value, all logical sectors become equally probable, so the probability of decoding success becomes the reciprocal of the number of logical sectors. This mapping not only provides an intuitive picture of error processes but also allows one to leverage established tools from statistical physics to analyze logical thresholds and failure rates \cite{PhysRevA.90.032326,SCHOLLWOCK201196,verstraete2004renormalizationalgorithmsquantummanybody}.

While the conceptual framework of topological QEC has matured considerably since its introduction, a quantitative understanding of logical failure rates remains a challenge. Precise error rate predictions depend on microscopic details such as lattice geometry, boundary conditions, physical noise channels, and the choice of decoder. This makes the analytic treatment of QEC codes operating in a practical regime an intractable problem in general. Focusing, instead, on the simpler post-selected limit of QEC codes can result in models that are tractable. In Ref.~\cite{english2024thresholdspostselectedquantumerror}, it was demonstrated that incorporating post-selection into the analysis of QEC codes constrains the quenched disorder in the corresponding statistical mechanics model, over which thermodynamic quantities are averaged. This constrained disorder viewpoint also underlies recent evidence that partial post-selection can yield scalable accuracy gains by discarding exponentially rare syndromes \cite{chen2025scalableaccuracygainspostselection}. Full post-selection is performed by aborting and reinitializing the code whenever any non-trivial syndrome is measured. In the extreme case of full post-selection on a surface code under a bit-flip error channel, the inherent disorder in the RBIM is entirely eliminated. This reduces the model to the clean two-dimensional Ising model, whose exact solution was first derived by Onsager in 1944 \cite{PhysRev.65.117}. This observation offers an analytic foothold to further our understanding of optimal QEC performance.

In this work, we exploit the mapping from a fully post-selected toric code under bit-flip noise to the two-dimensional Ising model, to derive closed-form analytic expressions for the logical failure rates of this code across its full domain of physical error rates and code distances. We identify four distinct regimes of topological error correction as a function of physical error rate and code distance. 

\begin{figure*}[thbp]
  \centering
  \subfloat[Mapping the toric code to an Ising model]{%
    \includegraphics[page=1,width=0.45\textwidth]{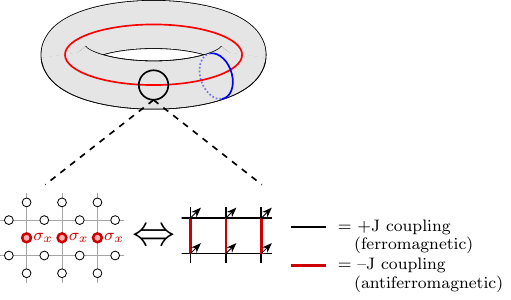}%
    \label{fig:errors_to_couplings}%
  }
  \subfloat[Four distinct regimes of error correction]{%
    \includegraphics[page=2,width=0.45\textwidth]{Figures.pdf}%
    \label{fig:error_regimes}%
  }\\
  \subfloat[Characteristics of the four regimes]{%
    \footnotesize
    \begin{tabular}{@{}l p{5.5cm} p{4.5cm} p{4.5cm}@{}}
      \toprule
      \textbf{Stat.~mech.~regime} & \textbf{Dominant physical mechanism} & \textbf{Logical failure behaviour} & \textbf{QEC interpretation} \\
      \midrule
      Path-counting & Combinatorics, minimum weight errors & $\sim p^{w_{\mathrm{min}}}$ & Performance set by minimal logical errors \\
      Ordered & Order, domain wall free energy cost & $\sim e^{-cd(\mathcal{L}_{m})}$ & Robust quantum memory, exponential error suppression \\
      Critical & Critical fluctuations, finite-size scaling & $\sim f((p-p_{c})L^{1/\nu})$ & Threshold behaviour, performance set by scaled distance to threshold \\
      Disordered & Disorder, proliferating defects & $\sim \mathbb{P}_{\mathrm{fail},\infty} - \mathcal{O}(e^{-c'd(\mathcal{L}_{m}^{C})})$ & Decoding fails reliably,  information lost \\
      \bottomrule
    \end{tabular}%
    \label{fig:regime_table}%
  }
\caption{\textbf{Decoding topological codes: A statistical mechanical perspective.} (a) The action of a logical operator in a topological code can often be mapped to the insertion of non-trivial defects or domain walls in an equivalent statistical mechanical model (e.g., $\pm J$ random-bond Ising model for certain surface codes, where $+J$ corresponds to ferromagnetic coupling between spins, and $-J$ corresponds to antiferromagnetic coupling). (b) The performance of optimal decoding exhibits distinct regimes as a function of physical error rate $p$ and code distance $L$, mirroring the regimes of the associated statistical mechanical model: Path-Counting, Ordered, Critical, and Disordered. The boundaries shown are schematic and may overlap at finite $L$: the Path-Counting and Critical regions denote finite-size crossover windows in which the corresponding approximations are expected to be accurate, whereas only the Ordered and Disordered regions remain as distinct thermodynamic phases in the limit $L\to\infty$. (c) Summary table providing the physical origin, functional behaviour of logical errors, and consequences for quantum error correction within each identified regime. These features are generic to topological codes whose decoding maps to classical statistical mechanics models that undergo second-order phase transitions. In the ordered phase, the table entry $\sim e^{-cd(\mathcal{L}_{m})}$ denotes the asymptotic suppression of the logical failure probability, where $d(\mathcal{L}_{m})$ is the support size of the minimal representative in the logical class $\mathcal{L}_{m}$. In the disordered phase, $\sim \mathbb{P}_{\mathrm{fail},\infty} - \mathcal{O}(e^{-c'd(\mathcal{L}_{m}^{C})})$ denotes exponential approach to the asymptotic failure probability $\mathbb{P}_{\mathrm{fail},\infty}$, with rate controlled by the support size $d(\mathcal{L}_{m}^{C})$ of the minimal representative in the conjugate logical class. In general, $d(\mathcal{L}_{m})$ and $d(\mathcal{L}_{m}^{C})$ need not coincide, although for the 2D toric/planar code both scale as $\mathcal{O}(L)$. For bit-flip noise on a single logical degree of freedom, $\mathbb{P}_{\mathrm{fail},\infty}=\frac{1}{2}$.
}
\label{fig:combined_figure}
\end{figure*}

First, at sufficiently low physical error rates, the system is well described by an approximation which becomes exact in the limit as the physical error rate $p$ goes to $0$. We call this the path-counting regime, in which failure rates can be estimated by counting the lowest-weight errors that cannot be corrected. Second, for sufficiently large code distances (or equivalently, lattice sizes), and for error rates below the topological code threshold, the system enters the below-threshold (ordered) regime, characterized by an exponential decay of logical failure rates with increasing code distance. Third, when the correlation length is comparable to the system size, finite-size scaling describes the critical region around the threshold $p_{c}$. Here, the scaling behaviour of logical failure rates is governed by the universality class of the phase transition in the underlying statistical mechanical model. Finally, for large code distances and physical error rates above the threshold, the system transitions into the above-threshold (disordered) regime, where the logical failure rate converges to the value dictated by a uniform distribution over the logical sectors. We use the term regime rather than phase to emphasize that these behaviours are defined for finite lattice sizes; in the thermodynamic limit, only the ordered and disordered phases remain. Fig.~\ref{fig:combined_figure} provides (a) a schematic overview of this mapping and (b) summarizes the resulting regimes along with (c) their physical interpretations.

The exact analytic expressions that have been derived for the two-dimensional Ising model allow a detailed analytical investigation of each of these regimes. For the fully post-selected toric code under bit-flip noise, the functional behaviours of logical failure rates throughout the different regimes are established exactly. Furthermore, we conjecture that while the introduction of disorder (or the absence of post-selection) may quantitatively shift certain parameters associated with the functional behaviour of logical failure rates in these regimes, the qualitative nature of the regimes remains robust across not only the non-post-selected surface code, but a broad class of other topological codes as well.
Here, we focus on $2D$ topological codes for which the noise models map to a classical statistical mechanical model that undergoes a second-order phase transition.
This work thus lays a rigorous foundation for understanding topological quantum error correction in finite, practical regimes -- bridging the gap between idealized, thermodynamic-limit descriptions and realistic, experimentally accessible systems.

The remainder of the manuscript is structured as follows. In Section \ref{sec:failure_free_energy}, we summarize the mapping between stabilizer codes and classical spin models. We explain how logical failure probabilities are encoded as free energy costs of domain walls and how these map onto decoding strategies, and we summarize key results from the spin glass literature on the scaling of domain wall energies. 
In Section \ref{sec:postselect}, we specialize to the fully post-selected toric code, exploiting its exact mapping to the clean 2D Ising model to derive closed-form expressions for the logical failure rate in each regime. In Section \ref{sec:no_postselect}, we extend these insights to non-post-selected surface codes, proposing effective tension and error function ans\"{a}tze that capture finite-size scaling and below-threshold behaviour. In Section~\ref{sec:circuit-level}, we investigate how the results developed in the preceding sections can be applied to circuit-level noise. Finally, in Section \ref{sec:discuss}, we extend the results from earlier sections into a general framework for topological QEC, outline its assumptions and scope, interpret the physical meaning of key parameters, and discuss limitations and avenues for future work.

\section{Failure rates and free energy}\label{sec:failure_free_energy}

In this section, we briefly summarize the formalism and key results from the literature on statistical mechanical mappings of QEC codes that underpin our later analyses. We assume familiarity with stabilizer codes, logical sectors and cosets, and the basic principles of decoding \cite{Nielsen_2012, gottesman1997stabilizercodesquantumerror}. We introduce the generalized statistical mechanical mapping of stabilizer codes \cite{10.1063/1.1499754,Chubb_2021}, the construction of the disordered Hamiltonian from physical error processes, and the correspondence between partition function calculations and decoding strategies \cite{PhysRevA.90.032326}. Where helpful, we summarize the distinctions between maximum-likelihood and maximum-probability decoding, as well as the role of post-selection in modifying the underlying spin model \cite{english2024thresholdspostselectedquantumerror}. We finally introduce key results from the spin glass literature, particularly the scaling of domain wall energy costs in random-bond Ising models \cite{PhysRevB.29.4026,PhysRevB.38.386}, which we leverage in our analyses below. 

Pauli stabilizer codes admit an exact mapping to disordered classical spin models. Each realization of the noise model for the code corresponds to a random choice of couplings in the classical model. The partition function of this model corresponds to a sum over error configurations that have the same logical action and cannot be distinguished by the stabilizers of the code. Through this mapping, the computation of disordered partition functions, free energy costs and phase transition points of the classical spin models directly yields logical failure probabilities and decoding thresholds \cite{10.1063/1.1499754,Chubb_2021}. We now summarize this correspondence, restricting our attention to qubit CSS codes, for pedagogical clarity. The stabilizer generators of such codes can be chosen to be either $X$-type or $Z$-type, containing only $\sigma_{X}$ and identity operators or only $\sigma_{Z}$ and identity operators, respectively. For a treatment beyond these restrictions, including general qudit stabilizer codes and correlated noise models, see Ref.~\cite{Chubb_2021}. In this mapping, a classical spin degree of freedom $s_{k}\in \mathbb{Z}_{2}$ is assigned to each stabilizer generator $\hat{S}_{k}$. Assuming single-qubit Pauli noise, spins interact if their corresponding stabilizers overlap on at least one physical qubit. 
Bit-flip noise induces interactions among spins associated with $X$-type stabilizers, phase-flip noise induces interactions among spins associated with $Z$-type stabilizers, and $Y$-type noise induces interactions involving both types of spins.
The coupling strength of each interaction is set by the underlying noise process acting on the corresponding qubit. In the absence of errors, all coupling strengths favour alignment of spins, which we refer to as ferromagnetic coupling. A single Pauli error on a physical qubit flips the sign of the couplings associated with that error, thereby introducing disorder into the system.

The statistical mechanical Hamiltonian produced by the mapping, for a given configuration of errors $\mathcal{E}$, is 
\begin{equation}\label{eq:Hamiltonian}
    H_{\mathcal{E}}(\vec{s})=-\sum_{i,\sigma\in\mathcal{P}_{i}}\overbrace{J_{i}(\sigma)}^\text{Strength}\overbrace{\llbracket\sigma,\mathcal{E}\rrbracket}^\text{Disorder}\overbrace{\prod_{k:\llbracket\sigma,S_{k}\rrbracket=-1}s_{k}}^\text{Interactions},
    \end{equation}
where $J_{i}(\sigma)$ are noise-dependent interaction strengths, $\mathcal{P}_{i}$ is the local Pauli group on qubit~$i$, the scalar commutator $\llbracket \sigma,\mathcal{E}\rrbracket=\pm 1$ distinguishes ferromagnetic from antiferromagnetic couplings, and $\mathcal{E}\in\mathcal{P}^{\otimes n}$ is a Pauli error, which defines the disorder on the system. We call this \textit{quenched disorder}, as it does not evolve with time. Consequently, thermodynamic quantities are averaged over the quenched disorder by first fixing the disorder realization, taking thermal averages, and then averaging this quantity over the quenched disorder distribution that is determined by the noise process on the physical qubits.

The following symmetry of the model relates the Hamiltonians corresponding to instances of the noise that are related by a stabilizer: 
$$H_{\mathcal{E}S_{k}}(\vec{s})=H_{\mathcal{E}}(\vec{s}+\hat{k}),$$
where $\hat{k}$ denotes flipping the $k^{\mathrm{th}}$ spin. In other words, flippping $s_{k}$ has the same effect as multiplying the physical error $\mathcal{E}$ by the stabilizer $\hat{S}_{k}$. Consequently, fixing $\mathcal{E}$ then summing over all spin configurations $\vec{s}$ is equivalent to summing over every error in the coset $\mathcal{E}\cdot \mathcal{S}$, i.e., every product of $\mathcal{E}$ with elements of the stabilizer group. Since the optimal MLD must consider all errors equivalent up to stabilizers, this symmetry makes the calculation of the partition function, where one sums over all spin configurations, a direct implementation of optimal decoding if the coupling strengths are chosen correctly.

We choose the coupling strengths of the model so that $e^{-\beta H_{\mathcal{E}}(\vec{0})}=\mathbb{P}(\mathcal{E})$ by construction. This is done so that by enumerating all possible spin configurations, we have $Z_{\mathcal{E}}=\sum_{\{\vec{s}\}}e^{-\beta H_{\mathcal{E}}(\vec{s})}\propto\mathbb{P}(\overline{\mathcal{E}})$. That is, the disordered partition function is proportional to the probability of the logical coset of which $\mathcal{E}$ is a representative error \footnote{The constant of proportionality reflects that codes which admit materialized symmetries, that is, non-empty subsets of stabilizer generators whose product is the identity induces gauge degrees of freedom in the spins which overcounts physical errors with a multiplicity equivalent in each logical coset \cite{Kitaev_2003}. That is, by flipping all spins of a given symmetry, we map back to the same physical error. These multiplicities cancel when taking fractions of partition functions.}. For these results to hold, it can be shown that the choice of the interaction strengths corresponds to the \textit{Nishimori conditions}, which is a manifold on which a gauge symmetry allows for several exact identities of thermodynamic quantities \cite{cdi_crossref_primary_10_1143_PTP_66_1169}. The Nishimori conditions are
\begin{equation}\label{eq:nishimori_conditions}
    \beta J_{i}(\sigma)=\frac{1}{|\mathcal{P}|}\sum_{\tau\in\mathcal{P}_{i}}\text{log}\mathbb{P}_{i}(\tau)\llbracket\sigma,\tau^{-1}\rrbracket,
\end{equation}
where $\beta$ is the thermodynamic beta of the model, $J_{i}(\sigma)$ are the coupling constants corresponding to the local Pauli $\sigma$ acting on qubit $i$, $\mathcal{P}_{i}$ are the local Pauli group acting on qubit $i$, $|\mathcal{P}|$ is the cardinality of the Pauli group considered (e.g., 2 for bit-flip, 4 for depolarizing) and $\mathbb{P}_{i}(\tau)$ is the probability of the local Pauli $\tau$ acting on qubit $i$.

With the Nishimori conditions satisfied and with disorder corresponding to a Pauli error $\mathcal{E}\in\mathcal{P}^{\otimes n}$, the free energy cost of a non-trivial logical operator $\mathcal{L}_{m}$ is
\begin{align}
    \Delta F_{m}(\mathcal{E}) &= F_{\mathcal{E}\mathcal{L}_{m}}(\beta_{N}) - F_{\mathcal{E}}(\beta_{N})\nonumber\\
    &= \frac{1}{\beta_{N}}\log\left(\frac{\mathbb{P}(\overline{\mathcal{E}})}{\mathbb{P}(\overline{\mathcal{E}\mathcal{L}_{m}})}\right),
\end{align}
where $\beta_{N}$ is the finite thermodynamic beta under which the Nishimori conditions are satisfied and $\mathbb{P}(\overline{\mathcal{E}})$ is the cumulative probability of all physical errors in the logical coset $\overline{\mathcal{E}}$. Successful decoding of a logical degree of freedom under MLD occurs when the free energy cost of a non-trivial logical operator, conditioned on an error $\mathcal{E}$ having occurred, is positive. Successful decoding of all logical degrees of freedom occurs when $\mathbb{P}(\overline{\mathcal{E}}) > \mathbb{P}(\overline{\mathcal{E}\mathcal{L}_{m}})$ for all non-trivial $\mathcal{L}_{m}$, conditioned on the error $\mathcal{E}$ having occurred. This optimal decoding corresponds to computing the free energy cost of domain walls along the Nishimori line at finite temperature, but is generically \#P-complete \cite{IyerPavithran2015HoDQ}. 

To overcome the computational intractability of MLD, one can also perform maximum-probability decoding (MPD). MPD selects the single most likely error in each logical coset without summing over stabilizer-equivalent degeneracies, and it can be expressed as the zero-temperature limit ($\beta\to\infty$) of the free energy cost calculation, where only the minimum-energy (or maximum-probability) ground state configuration remains. We write
\begin{align}
    \lim_{\beta\rightarrow\infty}\Delta F_{m}(\mathcal{E})&=\Delta E(\mathcal{E})
    \nonumber \\
    &= \min_{\vec{s}}H_{\mathcal{E}\mathcal{L}_{m}}(\vec{s})-\min_{\vec{s}}H_{\mathcal{E}}(\vec{s}).
\end{align}
Similarly, decoding success under MPD for this error $\mathcal{E}$ occurs if this ground state energy cost is positive.

As an example, a planar code with bit-flip noise maps to the 2D random-bond Ising model with $\pm J$ couplings. In such models, there is a single non-trivial logical operator $\mathcal{L}_{1}=\overline{X}$ which wraps around only one direction of the lattice. The ground state energy cost between logical sectors is $\Delta E(\mathcal{E})$, which we hereafter call $\Delta E$, leaving the functional dependence on the quenched disorder parameter implicit. $\Delta E$ is proportional to the difference in weights of the minimum-weight perfect matchings \cite{PhysRevB.75.174415}, which can alternatively be expressed as the signed logical gap as formulated in Refs.~\cite{gidney2023yokedsurfacecodes,PRXQuantum.5.010302,Smith_2024}. The logical failure rate under minimum-weight perfect matching decoding (MWPMD) can then be expressed as \cite{PRXQuantum.5.010302}
\begin{equation}
    \mathbb{P}_{\mathrm{fail}}=\int_{-\infty}^{0}d\Delta E\ \mathbb{P}(\Delta E).
\end{equation}

If the distribution of domain wall free energies of the disordered statistical mechanical model is known, then the optimal logical failure rate can be determined exactly for any code. As above, for surface codes with bit-flip noise, to compute the optimal logical failure rate, we integrate the probability distribution of the domain wall free energy cost over $\mathbb{R}^{-}$:
\begin{equation}
\mathbb{P}_{\mathrm{fail}}=\int_{-\infty}^{0}d\Delta F_{1}\ \mathbb{P}(\Delta F_{1}).
\end{equation}

Although the free energy cost of domain walls in the 2D random-bond Ising model has been extensively studied in the spin glass literature \cite{PhysRevB.79.184402,Fisch2006,PhysRevB.97.064410,PhysRevB.75.174415,PhysRevB.29.4026,PhysRevB.70.134425,PhysRevLett.91.087201,N.Kawashima_1997}, these insights remain underutilized in the quantum error correction community. For the $\pm J$ RBIM, two important phases are identified. In the ferromagnetic (ordered) phase ($p<p_{c}$), the system exhibits long-range ferromagnetic order, which means a non-zero spontaneous magnetization in the thermodynamic limit. However, individual bonds can still be frustrated, so the spins do not all align perfectly. The spin glass phase only occurs at $T=0$ in two dimensions \cite{PhysRevB.75.174415}, when $p>p_{c}$, i.e., above-threshold. In the spin glass (disordered) phase, frustration from the random bonds prevents simple long-range ferromagnetic order, leading to a frozen state with complex, spatially disordered spin configurations \cite{RevModPhys.58.801}. The critical point $p=p_{c}$ is the boundary between these two phases. Under certain noise models, some codes map to models that have been argued to undergo first-order phase transitions~\cite{Andrist_2011,rispler2024}. We nevertheless apply our approach when the order of the transition is uncertain and find a robust fitted threshold using our scaling ansatz in the examples we have studied. In such cases, the terms critical point and critical regime may not be accurate. We leave determining the order of these transitions to future work.

A key diagnostic in these spin glass descriptions is the distribution of the domain wall cost: the domain wall free energy cost $\Delta F$ along the Nishimori line $(\beta=\beta_N)$, relevant to optimal MLD, and the domain wall ground state energy cost $\Delta E$ in the zero-temperature limit $(\beta\to\infty)$, relevant to MPD. For a given disorder realization, these quantities measure the cost of inserting a nontrivial domain wall, and the logical failure rate is controlled by the probability that these domain walls lower the free energy of the system. Thus, the relevant object is not only the disorder-averaged cost, but the full distribution of $\Delta F$ or $\Delta E$.

Nevertheless, the scaling of the typical domain wall cost with system size provides an important diagnostic of the phase. The stiffness exponent $\theta$ is defined such that the rescaled domain wall cost, $\Delta F/L^{\theta}$ (or $\Delta E/L^{\theta}$ at $T=0$), has a nontrivial limiting distribution as $L\to\infty$ \cite{pandey2026}. In a ferromagnetic phase, $\theta=1$ as domain wall free energy costs are proportional to their length. In the $T=0$ spin glass phase of the 2D $\pm J$ RBIM, by contrast, $\theta=0$ \cite{PhysRevB.75.174415}, as the distribution $\mathbb{P}(\Delta F)$ converges to a limit independent of $L$ as $L\to\infty$ \cite{PhysRevB.97.064410}.

Between these limits lies a critical regime in which the typical domain wall cost grows sublinearly with $L$, and, more importantly for decoding, its full distribution exhibits nontrivial finite-size scaling. The distribution of domain wall costs in the RBIM with $\pm J$ disorder at $T=0$ remains understudied in the literature. However, for the RBIM with coupling constants drawn from a Gaussian distribution, along the low-temperature critical line below the multicritical point, it is known that $\Delta F\sim L^{\theta_{c}}$ with $\theta_{c}\approx 0.15$ \cite{PhysRevB.29.4026,PhysRevB.79.184402,pandey2026}. For the $\pm J$ RBIM, at the (finite-temperature) Nishimori multicritical point, it is also known that $\theta_{c}=0$, as this distribution is manifestly scale invariant \cite{wan2025}. It is this distributional scaling of $\Delta F$ and $\Delta E$ that we analyze in detail later in this work.

\section{Fully post-selected surface codes}\label{sec:postselect}

In Section \ref{sec:failure_free_energy}, we identified that the logical failure rate of QEC codes can be determined by the free energy costs of domain walls in the statistical mechanical model produced by an exact mapping. 
In previous works, these domain wall free energy costs have been considered as order parameters to determine a code's optimal threshold \cite{10.1063/1.1499754,PhysRevX.2.021004,Chubb_2021}. In this work, we go beyond simply estimating thresholds to obtain quantitive results for logical failure rates using the distribution of domain wall free energy costs. In this section, we map the fully post-selected toric code onto the clean 2D Ising model on a torus to obtain a closed-form expression for the logical failure rate across the full domain of physical error rates and code distances. We use this expression to analyze the functional behaviour of the logical failure rates of the code in the four regimes identified in Fig.~\ref{fig:combined_figure}. The behaviour of domain walls in these regimes is summarized in Fig.~\ref{fig:domain_walls}, which provides a guide for the quantitative analysis presented below.

\begin{figure}[thbp]
  \centering
  \subfloat[Path-counting]{%
    \includegraphics[width=0.2\textwidth]{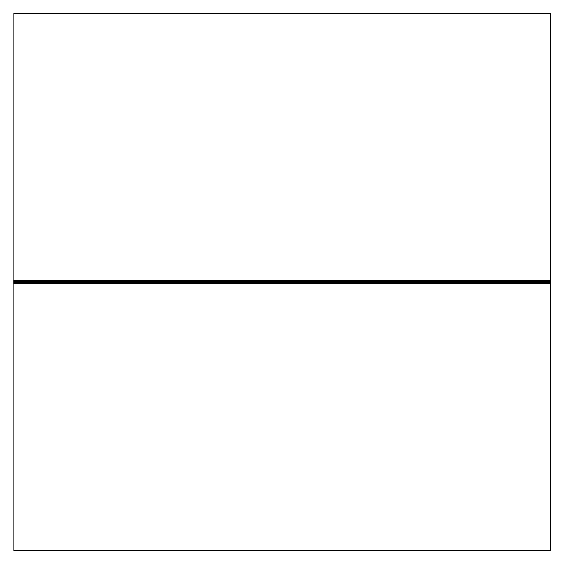}%
    \label{fig:path_counting_regime}%
  }
  \subfloat[Below-threshold]{%
    \includegraphics[width=0.2\textwidth]{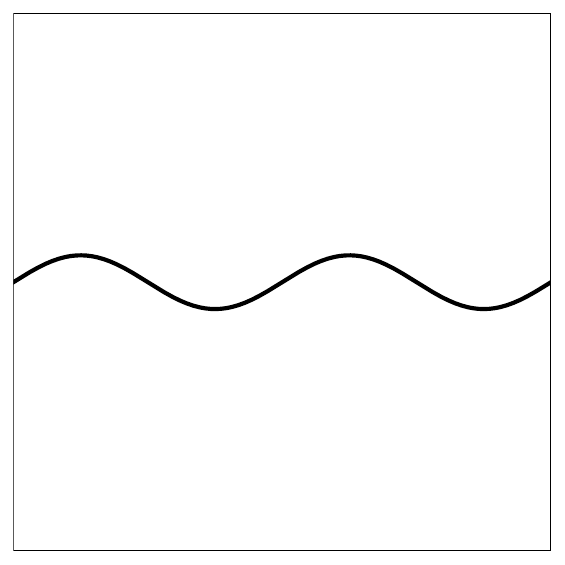}%
    \label{fig:ferromagnetic}%
  }\\
  \subfloat[Near-threshold]{%
      \includegraphics[width=0.2\textwidth]{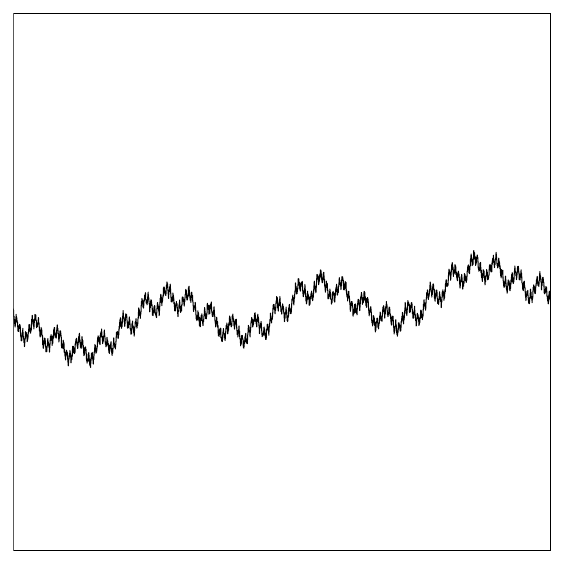}%
    \label{fig:scaling}%
  }
  \subfloat[Above-threshold]{%
      \includegraphics[width=0.2\textwidth]{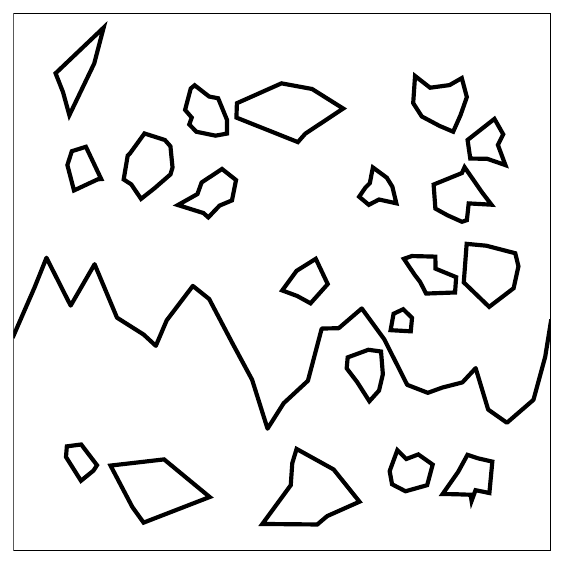}%
    \label{fig:paramagnetic}%
  }
\caption{\textbf{Domain wall behaviours across the four regimes of the fully post-selected toric code under bit-flip noise.}
(a) Path-counting regime ($p\ll p_{c}$): On a clean lattice at very low error rates, logical failure is dominated by the shortest possible domain walls, which remain essentially straight and unperturbed.
(b) Below-threshold regime ($p<p_{c}$): Below-threshold, thermal capillary wave fluctuations impart roughening to an otherwise predominantly straight interface, yielding small-amplitude corrections on a straight wall.
(c) Near-threshold regime ($p\approx p_{c}$): Near-threshold, scale invariance drives domain walls to develop fractal, self-similar structure, with fluctuations on all length scales reflecting critical behaviour.
(d) Above-threshold regime ($p>p_{c}$): Above-threshold, thermal fluctuations dominate, fragmenting the system into a dense tangle of short, disconnected domain wall loops.}
\label{fig:domain_walls}
\end{figure}

The 2D Ising model is one of the most thoroughly explored systems in physics, providing insights into critical phenomena and phase transitions. Originally solved exactly by Onsager in 1944 in the thermodynamic limit, the model exhibits a phase transition from a ferromagnetic (ordered) to a paramagnetic (disordered) phase at the critical point where $\beta_{c} J=\frac{1}{2}\log(1+\sqrt{2})$ \cite{PhysRev.65.117}. Onsager's exact solution marked a significant milestone in statistical mechanics which laid the groundwork for understanding universality and scaling behaviour near criticality.

The partition function of the Ising model on a finite torus with periodic-periodic $(pp)$ boundary conditions was first computed in 1949 by a map to the spin-representation of the group of rotations \cite{PhysRev.76.1232}. This computation was later generalized to antiperiodic-antiperiodic ($aa$), periodic-antiperiodic $(pa)$ and the converse ($ap$) using Grassmannian path integrals \cite{Ming-ChyaWu_2002} and a simplified transfer matrix method \cite{PhysRevE.66.057103}. We use these results for the different partition functions of the Ising model on the torus $Z_{pp},Z_{pa},Z_{ap},Z_{aa}$ to find exact analytic expressions of logical error probability of the toric code.

It was shown in Ref.~\cite{english2024thresholdspostselectedquantumerror} that the fully post-selected limit of a toric code under bit-flip noise maps directly to a clean two-dimensional Ising model. We will consider the case of a torus with $L$-by-$L$ lattice sites, although the results we describe can be extended to a rectangular lattice. The Nishimori conditions read
\begin{equation}\label{eq:toricbitflipNishimori}
\beta J = \frac{1}{2}\log\left(\frac{1-p}{p}\right),
\end{equation}
where $\beta$ is the thermodynamic inverse temperature of the clean 2D Ising model. For the clean 2D Ising model, the critical point occurs at $\beta_{c} J=\frac{1}{2}\log(1+\sqrt{2})$ \cite{PhysRev.65.117}. Substituting this into the Nishimori conditions gives the threshold error rate
\begin{equation}
p_c = \frac{1}{2+\sqrt{2}} \approx 0.2929.
\end{equation}

Under the general mapping of a quantum error correction problem to a disordered statistical mechanics model $Z_{\mathcal{E}}\propto\mathbb{P}(\overline{\mathcal{E}})$. The disordered partition function encodes the logical error probability summed over all possible errors consistent with the measured syndrome. In the fully post-selected case, we can describe the (unconditioned) probability of these logical errors through the Ising model's partition functions with different boundary conditions. Specifically, by fixing the index of logical operators with a boundary of the torus, we have
\begin{align}
    \mathbb{P}(\overline{\mathds{1}}\otimes\overline{\mathds{1}})&\propto Z_{pp},\nonumber\\
    \mathbb{P}(\overline{X}\otimes\overline{\mathds{1}})&\propto Z_{ap},\nonumber\\
    \mathbb{P}(\overline{\mathds{1}}\otimes\overline{X})&\propto Z_{pa},\nonumber\\
    \mathbb{P}(\overline{X}\otimes\overline{X})&\propto Z_{aa}.
\end{align}
The total logical failure probability is given by
\begin{equation}\label{eq:Pfail}
    \mathbb{P}_{\mathrm{fail}}=\frac{Z_{ap}+Z_{pa}+Z_{aa}}{Z_{pp}+Z_{ap}+Z_{pa}+Z_{aa}}.
\end{equation} Since we are interested in the square case we will have $Z_{pa}=Z_{ap}$.

Following Refs.~\cite{Ming-ChyaWu_2002,Izmailian_2012}, we evaluate the partition functions of the 2D Ising model for these boundary conditions in Appendix~\ref{app:fermions-on-torus}. The solution to the 2D Ising model on a lattice works by mapping the classical spins to a system of free fermionic variables. The fermions are massless at the phase transition where the model is described by a conformal field theory  \cite{DiFrancesco1997}. The fermions gain a mass $|\mu|$ away from criticality where
\begin{equation}\label{eq:mu}
    \mu=\log\left(\sqrt{\sinh(2J\beta)}\right),
\end{equation}
which is 0 at criticality as expected. The parameter $\mu$ is positive in the low temperature low $p$ regime and negative at high temperatures. This signed mass is useful for describing the physics on either side of the phase transition.

Along the two principal directions of the torus, the fermions may have periodic (Ramond) or antiperiodic (Neveu-Schwarz) boundary conditions \cite{DiFrancesco1997}, which leads to four sectors denoted by $(\alpha,\gamma)$, where $\alpha,\gamma\in\{0,\frac{1}{2}\}$. A value of $0$ (for $\alpha$ or $\gamma$) corresponds to periodic boundary conditions for the fermions in that respective direction, while a value of $1/2$ corresponds to antiperiodic conditions. The partition functions of the spin model, such as $Z_{pp}$, can be written in terms of partition functions of the fermions with these boundary conditions $Z_{0,0}, Z_{0,1/2},Z_{1/2,0},Z_{1/2,1/2}$. Specifically, the logical failure probability in terms of the fermion partition functions is
\begin{equation}
\label{eq:Pfailfermion}
\mathbb{P}_{\mathrm{fail}}=
    \frac{1}{2}+
    \frac{ \sgn(T-T_c)Z_{0,0}}{Z_{1/2,1/2} + Z_{0,1/2} + Z_{1/2,0} + \mathrm{sgn}Z_{0,0}}.
\end{equation}

Eq.~\eqref{eq:Pfailfermion} already provides significant information about the shape of the threshold plot for the fully post-selected code. At the critical point $p=p_c$ the solution gives $Z_{0,0}=0$ for all $L$ and thus $\mathbb{P}_{\mathrm{fail}}=1/2$. So the crossing of the logical error probabilities is perfect for all $L$ and occurs at $p=p_c$ and $\mathbb{P}_{\mathrm{fail}}=1/2$. In the limit of high $p$ and high $T$ all the partition functions are nearly equal and $\mathbb{P}_{\mathrm{fail}}\rightarrow 3/4$ as it should for the toric code with bit-flip noise.

\begin{figure}[t]
    \centering
    \includegraphics[width=\linewidth]{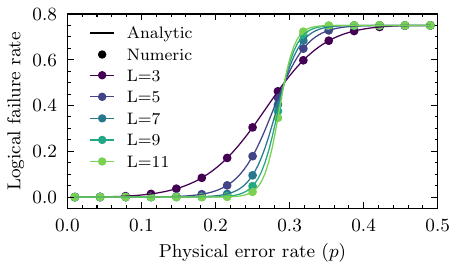}
    \caption{Logical failure rates for a fully post-selected toric code are shown as a function of the bit-flip probability $p$ for various lattice sizes $L$. The solid lines represent the exact analytic curves obtained via a mapping to the 2D Ising model, while the scatter points denote the failure rates computed using a tensor network decoder with a truncation bond dimension of $\chi=2^{8}$. The curves cross at the threshold error $p_c\simeq 0.29$ and logical failure probability $\mathbb{P}_{\mathrm{fail}}=1/2$ as predicted by the 2D Ising model solution.
}
    \label{fig:tn_vs_analytic}
\end{figure}

Using the explicit expressions for the $Z_{\alpha,\gamma}$ the probability of logical success of a fully post-selected toric code can be derived exactly. We show in Fig.~\ref{fig:tn_vs_analytic} the comparison of this analytic expression against the numerically simulated values taken from a tensor network decoder with bond dimension $\chi=2^{8}$ \cite{chubb2021generaltensornetworkdecoding}. As this tensor network decoder is optimal, without bond truncation, the numeric results are equal to the exact analytic expression. The curves all cross at the critical point as predicted by the discussion above.

This analytic expression for the logical failure rate of the fully post-selected toric code enables us to probe, in closed form, the functional behaviour of the failure rate across each of the four regimes identified above. Moreover, having this exact formula lets us explore failure rates for system sizes far larger than would be tractable via Monte Carlo sampling or splitting-method techniques \cite{PhysRevA.88.062308}. We now use this exact closed-form expression for the logical failure rate of a fully post-selected toric code to investigate the functional behaviour of these failure rates in each of the four regimes identified in Fig.~\ref{fig:combined_figure}.

\subsection{Path-counting regime}\label{subsec:path_counting_postselect}

For sufficiently low physical error rates the logical failure rate of a QEC code can be approximated by evaluating the total probability of all minimum-weight error configurations that will not be corrected successfully given some decoder. The regime in which these approximate expressions become accurate is known as the path-counting limit \cite{Watson_2014}, and in fact, in the limit as $p\rightarrow 0$, such approximations become exact. We obtain the path-counting expression for fully post-selected toric codes as follows. For sufficiently low  $p$, logical errors are dominated by a single domain wall forming that wraps around the torus. This probability without accounting for post-selection is
\begin{equation}
    \mathbb{P}_{\mathrm{wall}}=2Lp^{L}(1-p)^{2L^2-L}.
\end{equation}
The path-counting approximation is valid only as long as contributions from longer, higher-weight error chains are negligible. On the torus, the next-shortest error paths that cause a logical failure have a length of $L+2$. These paths can be visualized as the minimal, straight-line error chain deformed by two ``kinks" (see Appendix~\ref{app:path_count} for a detailed description). To assess when the length $L+2$ paths become significant, we compare their total probability to that of the minimal paths. The probability of any single path of length $L+2$ is suppressed by a factor of $(p/(1-p))^{2}$ relative to a minimal path. However, there is an additional combinatorial factor counting the number of ways such a path can be formed. The number of locations to place the two kinks on a chain of length $L$ scales as $L(L-1)$. The total contribution from the next-to-minimal paths is therefore smaller than the leading-order term by a factor of $L(L-1)(p/(1-p))^{2}$. For the path-counting approximation to hold, this term must be much less than 1, which implies the condition $p\ll 1/L$.

\begin{figure}[t]
    \centering
    \includegraphics[width=\linewidth]{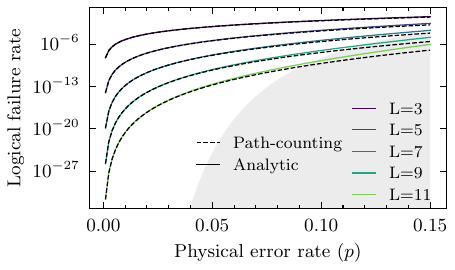}
    \caption{Logical failure rate $\mathbb{P}_{\mathrm{fail}}$ for a fully post-selected toric code under bit-flip noise in the path-counting regime. The failure rate is plotted as a function of the physical error rate $p$, with curves parametrized by the lattice size $L$. The dashed lines represent the path-counting approximation, while the solid lines show the exact analytic values. The shaded grey region indicates parameters where the path-counting regime is not expected to apply. Its boundary is obtained by substituting $L=1/p$ into the path-counting approximation (Eq.~\eqref{eq:path_count_postselect}), consistent with the validity scale $p\ll 1/L$ discussed in the main text.}
    \label{fig:path_counting}
\end{figure}

Next, we must use Bayes' rule to determine the probability of failure in the fully post-selected case. The event $\vec{s}=\vec{+1}$ includes all closed-loop error configurations, not just the empty configuration. However, the empty configuration gives the leading contribution, while the first corrections come from small contractible loops of length $4$, whose total weight is suppressed by $\mathcal{O}(L^{2}(p/(1-p))^{4})$. Since we are in the path-counting regime $p\ll 1/L$, these corrections are negligible, and hence
\[
\mathbb{P}(\vec{s}=\vec{+1}) \approx (1-p)^{2L^{2}}.
\]
We can then compute the path-counting expression for the logical failure rate in the fully post-selected regime using
\begin{align}\label{eq:path_count_postselect}
    \mathbb{P}_{\mathrm{fail}}&=\frac{\mathbb{P}_{\mathrm{wall}}}{\mathbb{P}(\vec{s}=\vec{+1})}\nonumber\\
    &\approx \frac{2Lp^{L}(1-p)^{2L^2-L}}{(1-p)^{2L^{2}}}\nonumber\\
    &= 2L\frac{p^{L}}{(1-p)^{L}}.
\end{align}
In Fig.~\ref{fig:path_counting}, we plot the predicted failure probability using the path-counting expression of Eq.~\eqref{eq:path_count_postselect} and the exact analytic values discussed above for a range of values of $L$. The approximation is good for small $p$ but once the physical error rate reaches a sufficiently large $p\sim 1/L$, the path-counting approximation breaks down and no longer captures the logical failure rate accurately. The path counting formula underestimates the logical error probability in this regime. At fixed code distance $L$, pushing $p$ above this scale drives the system into the below-threshold regime, which we now analyze.

\subsection{Below-threshold (ordered) regime}\label{subsec:below-threshold-postselect}

We have seen that the path counting regime holds only until $p\sim 1/L$. In this regime the expected number of errors in the whole system is only $\mathcal{O}(L)$. In the limit of large $L$ with $pL$ held constant the density of errors goes to zero as $1/L$. If instead we ask for non-zero error rate in the large $L$ limit we should hold $p<p_c$ constant as $L\rightarrow \infty$. Then we enter a regime where the system is macroscopic, with a constant non-zero density of errors. In this regime we can estimate the error probability $\mathbb{P}_{\mathrm{fail}}$ using known results about the free energy of domain walls in the 2D Ising model.

In the ordered regime of the Ising model the spins align as far as possible but the 3 different antiperiodic boundary conditions appearing in the formula for $\mathbb{P}_{\mathrm{fail}}$ require a single domain wall that will increase the free energy depending on the linear size of the system $L$. In the case of $Z_{ap}$ and $Z_{pa}$ these domain walls are vertical and horizontal respectively, and have length $L$. In the case of $Z_{aa}$ the domain wall runs diagonally and has length $2L$ (in the Manhattan metric). This will mean that $Z_{aa}$ is negligibly small compared to $Z_{ap}=Z_{pa}$.

We can define the change in free energy $\Delta F(\beta,L)$ due to the domain wall by the formula \begin{equation}\label{eq:domainwallfreeE}Z_{ap}=Z_{pa}=e^{-\beta\Delta F(\beta, L)}Z_{pp}.\end{equation}
In the thermodynamic limit, the free energy cost of a domain wall in the 2D Ising model is given by Onsager’s expression \cite{PhysRev.65.117},
\begin{equation}\label{eq:surfacetension}
    \Delta F(\beta,L\rightarrow \infty)=\sigma(\beta)L=\left(2J-\frac{1}{\beta}\log\coth(J\beta)\right)L,
\end{equation}
where $\sigma(\beta)$ is the domain wall surface tension. We work in units of $J=1$ when quoting $\sigma$ itself; however, the product $\beta\sigma$ is independent of this choice when the Nishimori conditions are satisfied, in which case $\beta \sigma$ can be written in terms of the error probability as
\begin{equation}\label{eq:cleanalpha}\beta \sigma(p)= \log\left(\frac{(1-p)(1-2p)}{p}\right).\end{equation}
See Appendix~\ref{app:asymptotics} for the asymptotic behaviour of \(\beta\sigma(p)\) as \(p\to0\) and \(p\to p_c\), which connects the path-counting and near-threshold regimes.

In finite systems, however, the interface is not perfectly rigid but undergoes capillary fluctuations~\cite{doi:10.1142/S0129183192000531,PhysRevE.90.012128}. These fluctuations modify the partition function by an amount that, after proper regularization, leads to a logarithmic correction to the free energy. Following an approximate capillary wave treatment (see Appendix \ref{app:capillary}), one finds that the finite-size correction leads to a domain wall free energy of the form
\begin{equation}\label{eq:freeEscaling}
    \Delta F(L,\beta)=\sigma(\beta)L-\frac{1}{2}\log(L)+\delta(\beta) + \mathcal{O}(1/L).
\end{equation}
The capillary wave correction is independent of $\beta$ in the below-threshold regime and the size-independent correction $\delta$ is known to have some temperature dependence, as seen, for example, in Fig.~9 of Ref.~\cite{PhysRevE.90.012128}. There are expected to be additional finite-size corrections of $\mathcal{O}(1/L)$ and smaller.

Neglecting the additive constant as a first approximation, we can substitute $\Delta F$ in Eq.~\eqref{eq:domainwallfreeE} and then evaluate the logical error probability to find
\begin{align}\label{eq:belowthreshold}
   \mathbb{P}_{\mathrm{fail}} & \simeq  \frac{Z_{ap}+Z_{pa}}{Z_{pp}} = 2e^{-\beta \Delta F(\beta,L)} \nonumber \\  & \simeq  2\sqrt{L} \left(\frac{p}{(1-p)(1-2p)}\right)^L +\mathcal{O}(p^L/\sqrt{L}).
\end{align}
In this approximation the contribution of $Z_{aa}$ is neglected since it is of order $p^{2L}$ and expected to be smaller than the subleading free energy corrections to $Z_{ap}$ that we have neglected.

We can compare the below-threshold logical failure rate (Eq.~\eqref{eq:belowthreshold}) to the expression that holds in the path counting regime (Eq.~\eqref{eq:path_count_postselect}). At low $p$ the dominant exponential suppression of errors approaches $p^n$ in both cases. The increased error rate at moderate $p$ is captured by the additional factor of $(1-2p)^{-L}$ in Eq.~\eqref{eq:belowthreshold}. Capillary waves cause the domain wall to wander over long length scales as indicated in Fig.~\ref{fig:domain_walls}. The factor of $\sqrt{L}$ in Eq.~\eqref{eq:belowthreshold} can be interpreted as describing the effective domain wall width that results. The ranges of validity of Eq.~\eqref{eq:belowthreshold} and Eq.~\eqref{eq:path_count_postselect} do not precisely overlap so that these logarithmic contributions to the free energy do not agree between the two limits we have used to find approximate expressions.

\begin{figure}[t]
    \centering
    \includegraphics[width=\linewidth]{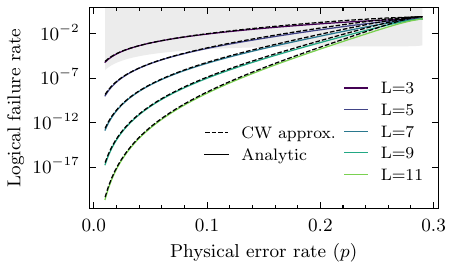}
    \caption{Comparison of the capillary wave approximation (dashed lines) and the exact analytic calculation (solid lines) of the logical failure rate for a fully post-selected toric code. Results are shown as a function of the physical error rate $p$ for various lattice sizes $L$, with each colour representing a different $L$. The grey shading marks parameters where the capillary-wave expression (Eq.~\eqref{eq:belowthreshold}) is not expected to be accurate. The boundary is obtained by imposing the near-threshold scale $\lvert p-p_c\rvert L=1$ (i.e., $\xi \sim L$; see Sec.~\ref{subsec:near-threshold-postselect}) and, on the below-threshold side relevant here, substituting $p=p_c-1/L$ into Eq.~\eqref{eq:belowthreshold}; we shade the region above this boundary.}
    \label{fig:sub-threshold}
\end{figure}

We plot in Fig.~\ref{fig:sub-threshold} comparison of this approximate expression with the exact analytically computed failure rates using the free fermion mapping.

Fig. \ref{fig:sub-threshold} shows that as $p$ approaches the critical value $p_{c}$, our capillary wave approximation breaks down because domain wall fluctuations grow too large. These fluctuations are growing since the surface tension decreases towards zero as the critical point is approached. Eq.~\eqref{eq:belowthreshold} is not expected to be valid when the system size $L$ becomes comparable to $\beta \sigma$ and the logical error probability is $\mathcal{O}(1)$. In the next section we turn to finite-size scaling theory to describe the logical failure rates in this critical regime.

\subsection{Near-threshold (critical) regime}\label{subsec:near-threshold-postselect}

As we approach the threshold error rate $p_{c}$, fluctuations span the entire lattice and finite-size effects dominate \cite{10.1093/oso/9780198517962.001.0001}. The fluctuations are described by a correlation length $\xi(\beta)$ that becomes very large in the region of the phase transition. For system sizes $L\leq\xi$ unitless physical quantities such as the logical failure probability $\mathbb{P}_{\mathrm{fail}}(\beta,L)$ are functions of $L/\xi$ only. For a general statistical mechanical model close to a critical point the correlation length $\xi$ diverges as $\xi\propto|\beta-\beta_{c}|^{-\nu}$ for some scaling exponent $\nu>0$. Comparing to the Nishimori condition~(Eq.~\eqref{eq:toricbitflipNishimori}) we see that we also have $\xi\propto |p-p_c|^{-\nu}$ in this regime. In the case of the 2D Ising model it is known that $\xi(\beta)=1/\beta\sigma(\beta)$~\cite{Mccoy_and_wu} and by differentiating $\sigma$ as given in Eq.~\eqref{eq:surfacetension} at $\beta_c$ it is straightforward to confirm the known result that $\nu=1$. 

\begin{figure}[t]
    \centering
    \includegraphics[width=\linewidth]{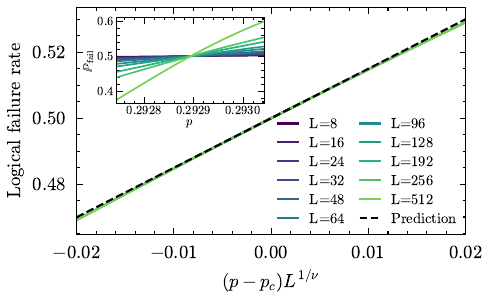}
    \caption{Data collapse of the logical failure rate $\mathbb{P}_{\mathrm{fail}}$ for the fully post-selected toric code under bit-flip noise. The logical failure rate is plotted against the scaling variable $x=(p-p_{c})L$, where $p_{c}=\frac{1}{2+\sqrt{2}}$ is the critical error rate. Near criticality, the data collapse onto a single curve consistent with the scaling form $\mathbb{P}_{\mathrm{fail}}(p,L)=f[(p-p_{c})L^{1/\nu}]$, with $\nu=1$ for the 2D Ising universality class. The dashed line shows the first-order prediction for the scaling function derived from the free fermion partition functions in Appendix~\ref{app:near_threshold_expansion}. \emph{Inset:} The data before rescaling.}
    \label{fig:data_collapse}
\end{figure}

We now define a dimensionless scaling variable
\[
x=(p-p_{c})L^{1/\nu}\simeq (L/\xi)^{1/\nu}.
\]
In the critical scaling regime we expect that
\begin{equation}\label{eq:genscaling}\mathbb{P}_{\mathrm{fail}}(p,L)=f[(p-p_{c})L^{1/\nu}].\end{equation}
If we plot the logical error probability against $x$ for all $L$ then the data collapses onto a single curve given by the scaling function $f(x)$. Since we have an analytical solution to the 2D Ising model it is possible to see exactly how the scaling limit emerges from the solution. In fact the scaling function $f$ could in principle be determined from the exact solution of the Ising model. We can already see from our earlier discussion that $f(0)=1/2$ and $f(x)\rightarrow 3/4$ as $x\rightarrow \infty$. By performing a Taylor expansion of $\mathbb{P}_{\mathrm{fail}}(x)$ centred at $x=0$, we can see that
\begin{equation}
    \mathbb{P}_{\mathrm{fail}}(x)\approx \frac{1}{2} + f'(0)x
\end{equation}
for small $x$, where the slope $f'(0)$ can be obtained through a first-order expansion (in the fermion mass $\mu$) of the partition functions with different boundary conditions. In Appendix~\ref{app:near_threshold_expansion}, we use the results of Ref.~\cite{Ivashkevich_2002} to perform this expansion, obtaining $f'(0)\approx1.503$. Fig.~\ref{fig:data_collapse} demonstrates this data collapse, particularly close to the phase transition, and the dashed black line indicates the derived first-order prediction of the scaling function $f$. Once $x \ll 0$ or $x \gg 0$ (i.e., $L\gg \xi$), the curves no longer fall onto a single, $L$-independent line. 

We now show explicitly how finite-size scaling behaviour arises from the low-momentum modes in the free fermion mapping in order to understand the regime of validity of the near-threshold regime. The fermion partition functions $Z_{\alpha,\gamma}(\mu,L)$ depend on the signed mass parameter $\mu$ and $L$ through the boundary conditions, which determine the allowed fermion momenta $k$, and $L\omega_\mu(k)$ where the lattice dispersion relation is
\begin{equation}\label{eq:dispersion}
    \omega_{\mu}(k)=\mathrm{arcsinh}\left(\sqrt{\sin^{2}(k) + 2\sinh^{2}(\mu)}\right).
\end{equation}
Near criticality, Eq.~\eqref{eq:mu} implies $\mu\simeq 0$. Since the scaling plot is written in terms of $p$, we expand $\mu$ about $p=p_c$:
\begin{equation}\label{eq:mu_expansion_fss}
    \mu\approx\left.\frac{d\mu}{dp}\right|_{p=p_c}(p-p_c).
\end{equation}
Using
\[
\mu=\log\left(\sqrt{\sinh(2J\beta)}\right),
\qquad
\beta J=\frac{1}{2}\log\left(\frac{1-p}{p}\right),
\]
we find
\[
\frac{d\mu}{d (\beta J)}=\coth(2\beta J),
\qquad
\frac{d(\beta J)}{dp}=-\frac{1}{2p(1-p)}.
\]
At the critical point $p_{c}=\frac{1}{2+\sqrt{2}}$, equivalently $\sinh(2\beta_{c} J)=1$, this gives
\[
\left.\frac{d\mu}{dp}\right|_{p=p_{c}}=-(2+\sqrt{2}).
\]
Hence
\begin{equation}\label{eq:mu_x_relation}
    \mu \approx -(2+\sqrt{2})(p-p_c) = -\frac{(2+\sqrt{2})x}{L}.
\end{equation}
Thus $|\mu|\sim \xi^{-1}$ in the critical regime, as expected for the inverse correlation length. Eq.~\eqref{eq:mu_x_relation}, together with Eq.~\eqref{eq:dispersion}, shows that the low energy sector is controlled by the scaling variable $x$.

For the low lying momentum modes relevant to finite-size scaling,
\[
k=\frac{\pi(n+\alpha)}{L},
\qquad n=0,1,\ldots,
\]
with $n$ fixed as $L\to\infty$. Since both $\mu$ and $k$ are small when $x=\mathcal{O}(1)$, we may expand $\sin(k)\approx k$ and $\sinh(\mu)\approx \mu$ in Eq.~\eqref{eq:dispersion}. This gives
\begin{align}\label{eq:Lomega}
    L\omega_{\mu}\left(\frac{\pi(n+\alpha)}{L}\right)
    &\approx
    \sqrt{\bigl(\pi(n+\alpha)\bigr)^2 + 2(\mu L)^2}
    \nonumber\\
    &=
    \sqrt{\bigl(\pi(n+\alpha)\bigr)^2 + 2(2+\sqrt{2})^2x^2}.
\end{align}
This quantity is $\mathcal{O}(1)$ for fixed $n$ and $x=\mathcal{O}(1)$.

Each of the fermion partition functions $Z_{\alpha,\gamma}$, as defined in Appendix~\ref{app:fermions-on-torus}, depends only on $L\omega_{\mu}(k)$ for each mode $k$. Thus the contribution to the partition functions $Z_{\alpha,\gamma}$ of the modes with $n$ less than some constant, say 100, are given by Eq.~\eqref{eq:Lomega} and only depend on $x$. 

As a result near criticality each full partition function can be decomposed into a universal part corresponding to these low values of $n$ that depend only on $x$ and non-universal contributions that arise from higher-energy modes with larger values of $n$. When forming the ratio in Eq.~\eqref{eq:Pfailfermion}, these non-universal factors appear in every term in a similar way, so they cancel out. This cancellation leaves the logical failure probability dependent solely on the scaling variable $x=(p-p_{c})L$, capturing the universal critical behaviour. 

The emergence of this universal scaling, derived in this $x=\mathcal{O}(1)$ regime for the fully post-selected case (see Fig.~\ref{fig:data_collapse}), is particularly pronounced within a characteristic window where the scaling variable x is of order unity (e.g., $|x|<c$ for some constant $c$). This observation provides the foundation for expecting similar finite-size scaling behaviour in the non-post-selected toric code, albeit with a scaling variable $x=(p-p_{c})L^{1/\nu}$, with $\nu\neq 1$, reflecting its distinct universality class.

At fixed code distance $L$, increasing the error rate $p$ beyond $p_{c}$ drives the system into the above-threshold regime, where domain walls proliferate and the code loses its ability to protect quantum information. This is the topic of the next subsection.

\subsection{Above-threshold (disordered) regime}

The above-threshold regime occurs at higher bit-flip probabilities, i.e., above the error threshold. In this regime, the dominant thermal fluctuations disrupt long-range order, leading to fundamentally different behaviour in the Ising model and, consequently, in the logical failure rate of the post-selected toric code. We use a Kramers-Wannier duality to map the behaviour of the below-threshold regime to the above-threshold regime \cite{PhysRev.60.252}. This shows that the logical failure probability $\mathbb{P}_{\mathrm{fail}}$ exponentially decays from unity toward $3/4$ with increasing $L$, consistent with the fourfold degeneracy of logical cosets in the high-error limit.

The dual thermodynamic beta is defined through \cite{baxter2016exactly}
\begin{equation}
    \sinh(2\beta J)\sinh(2\beta^{*}J)=1,
\end{equation}
or more conveniently,
\begin{equation}
    \tanh(J\beta^{*})=e^{-2J\beta}.
\end{equation}
Substituting in the Nishimori conditions, we get
\begin{align}\label{eq:tanhJB1}
    \tanh(J\beta^{*})=\frac{p}{1-p}.
\end{align}
We can similarly write 
\begin{align}\label{eq:tanhJB2}
    \tanh(J\beta^{*})&=\frac{1-e^{-2J\beta^{*}}}{1+e^{-2J\beta^{*}}}\nonumber\\
    &= \frac{1-\frac{p^{*}}{1-p^{*}}}{1+\frac{p^{*}}{1-p^{*}}}\nonumber\\
    &= 1-2p^{*},
\end{align}
and equating Eq.~\eqref{eq:tanhJB1} and \eqref{eq:tanhJB2}, we obtain
\begin{equation}
    p^{*}=\frac{1-2p}{2p(1-p)}.
\end{equation}

The Kramers-Wannier duality mixes the different sectors of the partition functions as \cite{PhysRevB.55.11045}
\begin{equation}
    \begin{bmatrix}
           Z_{pp}(p) \\
           Z_{pa}(p) \\
           Z_{ap}(p) \\
           Z_{aa}(p)
         \end{bmatrix} = \frac{1}{2} \begin{bmatrix}
             1 & 1 & 1 & 1\\
             1 & 1 & -1 & -1 \\
             1 & -1 & 1 & -1 \\
             1 & -1 & -1 & 1
         \end{bmatrix} \begin{bmatrix}
           Z_{pp}(p^{*}) \\
           Z_{pa}(p^{*}) \\
           Z_{ap}(p^{*}) \\
           Z_{aa}(p^{*})
         \end{bmatrix}.
\end{equation}
So immediately it follows that
\begin{align}\label{eq:KW_dual}
    [1-\mathbb{P}_{\mathrm{fail}}(p)][1-\mathbb{P}_{\mathrm{fail}}(p^{*})]&=\frac{1}{4},
\end{align}
We can therefore estimate the logical failure rates in the above-threshold regime using the capillary wave theory derived in the below-threshold regime. In the below-threshold regime, we showed that the failure rate exponentially decayed to $0$ as $L\to\infty$ for fixed $p<p_{c}$. Then by Eq.~\eqref{eq:KW_dual}, we see that our logical failure rate in the above-threshold regime exponentially approaches $3/4$ at fixed $p>p_{c}$.

In the next section, we move beyond the fully post-selected scenario. Before moving on, we note that the analytical expressions for the fully post-selected surface code derived in this section apply directly to various magic-state injection, distillation, and cultivation schemes (e.g., Refs.~\cite{PRXQuantum.5.010302,2409.17595,lee2025lowoverheadmagicstatedistillation}).

\section{Surface Codes Without Post-selection}\label{sec:no_postselect}

In this section, we consider the standard setting of quantum error correction without post-selection. We thereby introduce quenched disorder into the statistical mechanical model. It is well known that the toric code under the bit-flip error model can be mapped to the random bond Ising model in 2D (RBIM) \cite{10.1063/1.1499754}. Although this model is not exactly solved like the clean Ising model, there is a large literature on this model due to its relevance to spin glasses \cite{PhysRevB.79.184402,Fisch2006,PhysRevB.97.064410,PhysRevB.75.174415,PhysRevB.29.4026,PhysRevB.70.134425,PhysRevLett.91.087201,N.Kawashima_1997}.

Here we investigate two complementary decoding settings for the non-post-selected surface code: the zero-temperature limit $T=0$, corresponding to MWPMD, and along the Nishimori line $T=T_{N}(p)$, corresponding to optimal MLD. In the statistical mechanical picture, these probe, respectively, domain wall ground state energy costs and domain wall free energy costs in the disordered model. Comparing these two settings allows us to distinguish features that are specific to the full finite temperature decoding problem from those already present in the ground state structure of the disordered model. It also allows us to compare the behaviour of a standard efficient decoder with that of the optimal decoder, while retaining a common statistical mechanical interpretation.

For the $T=0$ numerics, we use MWPMD as implemented in \emph{PyMatching} \cite{pymatchingv2}. We note, however, that the zero-temperature decoding problem is highly degenerate: a given syndrome may admit many minimum-weight corrections, and a specific matching implementation does not in general fairly sample this degenerate manifold \cite{PhysRevA.81.022317}. As emphasized in Ref.~\cite{PhysRevA.81.022317}, this can affect precise quantitative estimates of zero-temperature threshold properties. Accordingly, our $T=0$ results should be interpreted primarily as characterizing the phenomenology of this standard matching decoder and the associated regime structure, rather than as an unbiased high-precision determination of the zero-temperature RBIM critical point.

For the finite $T=T_{N}(p)$ numerics, we use the \emph{Planar} decoder \cite{PhysRevLett.134.190603}, which uses the Kac-Ward determinant formula to compute partition functions of planar disordered Ising spin glasses, enabling exact evaluation of the free energy costs of domain walls for fixed disorder realizations. The Kac-Ward determinant formula for computing such partition functions can be shown to be equivalent to the FKT algorithm \cite{PhysRevE.87.043303,Cimasoni_2010}.

We demonstrate that the four regimes and their qualitative failure rate behaviours, identified in Section \ref{sec:postselect} for the post-selected case, remain robust even for non-post-selected surface codes. This is underpinned by the expectation that the four regimes identified from the post-selected surface code carry across to the random-bond Ising model that describes the system with quenched disorder, corresponding to error correction with no post-selection. We now detail the behaviour of surface codes within these four regimes. For each regime, we describe its key characteristics and introduce the specific ansatz for the logical failure rate $\mathbb{P}_{\mathrm{fail}}(p,L)$ that we will derive and analyze in the subsequent discussion:

\begin{enumerate}
    \item \textbf{Path-counting regime}: At sufficiently low $p\ll 1/L^{2}$, one can determine the probability of errors occurring that have the minimum Hamming weight required to inflict a logical error upon decoding. In this regime, as detailed in Sec.~\ref{subsec:path-count_nopostselect}, we show that the failure rate for a toric code under the bit-flip error model with odd $L$ follows the form
    \[
    \mathbb{P}_{\mathrm{fail}}(p,L)\approx\frac{2L\ L!}{\lceil L/2 \rceil!\lfloor L/2 \rfloor!}p^{\lceil L/2 \rceil}.
    \]
    
    \item \textbf{Below-threshold regime} (ordered): This regime occurs for $p<p_{c}$ and sufficiently large $L\gg 1$, and is identified with logical failure rates that decay exponentially to $0$ with code distance. In this regime, logical failure is controlled by the negative tail of the domain wall free energy distribution. As will be shown in Sec.~\ref{subsec:subthreshold_nopostselect}, for the toric code under the bit-flip error model we expect, over the numerically accessible finite-size regime, the failure rate of each logical degree of freedom to follow the effective form
    \[
    \mathbb{P}_{\mathrm{fail}}(p,L)\approx e^{-\beta[\sigma_{\mathrm{eff}}(p)L+\delta(p)]},
    \]
    where $\sigma_{\mathrm{eff}}(p)$ is an \emph{effective tension}, and $\delta(p)$ is a finite-size offset that absorbs subleading corrections. Within this finite-size parametrization, this offset sets the $\mathcal{O}(1)$ failure probability when the fitted $\sigma_{\mathrm{eff}}$ vanishes.

    \item \textbf{Near-threshold regime} (critical): Near criticality, low-momentum modes dominate the physics, and finite-size scaling implies that the logical failure rate depends on the scaling variable $x=(p-p_c)L^{1/\nu}$, equivalently on the ratio $L/\xi$, where $\xi\sim |p-p_c|^{-\nu}$. For the clean fully post-selected problem, the transition lies in the 2D Ising universality class. For the disordered RBIM relevant to non-post-selected decoding, however, the relation to the clean Ising fixed point is more subtle. Because the 2D Ising model is marginal under the Harris criterion \cite{ABHarris_1974}, weak quenched bond randomness is \emph{marginally irrelevant}: along the paramagnetic-ferromagnetic transition line near the pure Ising point, the critical behaviour is still controlled by the pure Ising fixed point, with logarithmic corrections rather than changed critical exponents \cite{ParisenToldin2009,PhysRevE.78.011110}.

    At sufficiently strong disorder, the phase diagram crosses over to distinct strong-disorder fixed points. In the 2D $\pm J$ RBIM, the Nishimori line is RG-invariant \cite{PhysRevB.40.9249,PhysRevLett.61.625} and intersects the paramagnetic-ferromagnetic boundary at the Nishimori multicritical point \cite{PhysRevB.63.104422,PhysRevB.55.1025,cdi_crossref_primary_10_1143_PTP_66_1169}, which is the point relevant to maximum-likelihood decoding \cite{10.1063/1.1499754,Chubb_2021}. Along the Nishimori line itself -- the direction relevant when varying the physical error rate in the decoding problem -- numerical work finds $\nu\approx 1.5$ \cite{PhysRevB.65.054425,ParisenToldin2009,wan2025}, and recent analytical work argues for the exact value $\nu=3/2$ \cite{Delfino_2025}.

    Below the Nishimori multicritical point, the low-temperature paramagnetic-ferromagnetic boundary is believed to be controlled by a strong-disorder fixed point distinct from the clean Ising one, with critical behaviour numerically consistent with that of the $T=0$ endpoint \cite{ParisenToldin2009}. This low-temperature strong-disorder universality class is believed to be distinct from the Nishimori multicritical point, although the correlation length critical exponent at the $T=0$ ferromagnet-spin glass transition has also been found numerically to be close to $\nu\approx 1.5$ \cite{WangChenyang2003Ctia,Watson_2014}.
    
    Motivated by this picture, we show in Sec.~\ref{subsec:nearthresh_nopostselect}, by analyzing the distribution of domain wall energy costs, that the failure rate for the surface code with bit-flip noise in this regime is governed by the finite-size scaling ansatz
    \[
    \mathbb{P}_{\mathrm{fail}}(p,L)\approx \frac{1}{2}\biggl[1-\mathrm{erf}\left(\frac{A_{2}x^{2}+A_{1}x+A_{0}}{\sqrt{2}}\right)\biggr],
    \]
    where $x=\mathrm{sgn}(p-p_c)(L/\xi)^{1/\nu}=(p-p_{c})L^{1/\nu}$ is the scaling variable, and $A_{2},A_{1}$ and $A_{0}$ are constants that relate to the domain wall energy cost distribution at criticality.
    
    \item \textbf{Above-threshold regime} (disordered): This regime occurs for $p>p_{c}$ and sufficiently large $L\gg 1$, and is identified with logical failure rates which exponentially approach $(K-1)/K$ with code distance: \[\mathbb{P}_{\mathrm{fail}}(p,L)\approx 1-\frac{1}{K}e^{-\beta[\sigma'_{\mathrm{eff}}(p)L+\delta'(p)]}.\] The introduction of quenched randomness fundamentally alters the Kramers-Wannier duality. Unlike the pure model, the duality does not provide a simple mapping between the RBIM at parameters $p$ and a dual point $p^{*}$ \cite{PhysRevE.107.024125}, and consequently, the elegant transformation relating the above-threshold behaviour to the below-threshold regime is lost. As a result we do not have a way of inferring either $\sigma'$ or $\delta'$ from the below threshold values $\sigma$ and $\delta$. Due to the complexities arising from quenched disorder, particularly the loss of a simple duality, developing and verifying a more precise ansatz for the approach to the asymptotic failure rate in this regime is beyond the scope of the current work.
\end{enumerate}

Having outlined the expected characteristics of the four operational regimes in surface codes without post-selection, we now aim to quantitatively model their failure rates. The exact analytical solutions for the post-selected system, presented in Section \ref{sec:postselect}, provide a foundation for developing suitable ans\"{a}tze. The subsequent analysis, presented for each regime, serves to test these models and thereby substantiate the robustness of the four-regime structure in the presence of quenched disorder.

\subsection{Path-counting}\label{subsec:path-count_nopostselect}

In this subsection, we restrict attention to the $T=0$ MWPMD data. The path-counting regime is most naturally formulated in this zero-temperature setting, since it describes the asymptotic $p\to 0$ limit, which on the Nishimori line also implies $T_{N}(p)\to 0$. The derivation of the logical failure rate expression in the path-counting regime is similar to the post-selected case. One assumes that logical failure is dominated by the minimal uncorrectable error configurations, which without post-selection are those in which exactly $\lceil L/2 \rceil$ errors occur along a domain wall of length $L$ \cite{PhysRevA.86.032324,Watson_2014,Beverland_2019,10.1063/1.1499754}. Under the assumption that errors are independent and occur with probability $p$, the probability for a given minimal path is proportional to $p^{\lceil L/2 \rceil}$. For a toric code with an odd distance $L$, including the combinatorial multiplicity of such paths leads to an expression of the form
\begin{equation}\label{eq:path_count}
    \mathbb{P}_{\mathrm{fail}}(p,L)=\frac{2L\ L!}{\lceil L/2 \rceil!\lfloor L/2 \rfloor!}p^{\lceil L/2 \rceil},
\end{equation}
where the $L$-dependent prefactor accounts for the number of distinct domain wall configurations. For an even distance $L$, the minimal error chains of weight $L/2$ cause a decoder failure only half the time, which introduces an additional factor of 1/2 to the prefactor.

This expression represents the leading-order contribution to the logical failure rate, becoming increasingly accurate as $p\to 0$, where such minimal error configurations dominate all other failure mechanisms. The subsequent terms in this expansion, which depend on longer error paths, determine the range of $p$ for which this leading-order approximation is valid. Following the approach of Dennis \emph{et al.}~\cite{10.1063/1.1499754} and Beverland \emph{et al.}~\cite{Beverland_2019}, we present a detailed calculation of the next leading-order contributions in Appendix~\ref{app:path_count}. This allows us to estimate the regime of validity for the path-counting estimate, which we find to be $p\ll 1/L^2$, and highlights a strong dependence on boundary conditions so that the toric code, the planar code with even and odd $L$, and the rotated planar code all have different error rates. This very strong boundary condition dependence is not observed for large $L$. We refer to Beverland \emph{et al.}~\cite{Beverland_2019} for a more complete discussion of these points.

\begin{figure}[h]
    \centering
    \includegraphics[width=\linewidth]{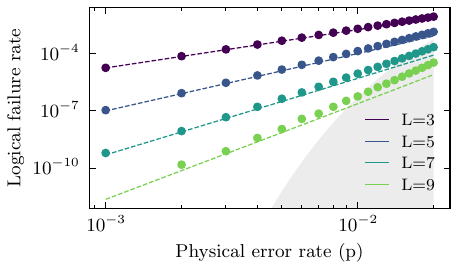}
    \caption{Comparison of the logical failure rate of a standard toric code (without post-selection) in the path-counting regime. Numerical simulation results (scatter points) are shown alongside the theoretical prediction from the path-counting approach (dashed lines). Data points represent between $n=10^{9}$ and $n=6\times10^{10}$ samples, with decoding performed using \emph{PyMatching} \cite{pymatchingv2}. The agreement improves significantly as $p$ decreases, confirming the accuracy of the path-counting expression in the low physical error rate limit. The grey shaded region indicates parameters where the path-counting expression is not expected to be accurate. The boundary is obtained by imposing $p=1/L^{2}$ (see the validity scale $p\ll1/L^{2}$ discussed in the main text) in the path-counting expression (Eq.~\eqref{eq:path_count}), and we shade to the right (larger $p$) of this boundary. For a smooth boundary across non-integer $L$, we evaluate a continuous $\Gamma$-function extension of the combinatorial prefactor in Eq.~\eqref{eq:path_count}, which agrees with the integer expression at integer $L$ (up to even/odd rounding) and is used only for the boundary visualization.}
    \label{fig:path_count_accuracy}
\end{figure}

To explore the validity of the path-counting approach, we performed numerical simulations at very low physical error rates. As shown in Fig.~\ref{fig:path_count_accuracy}, the path-counting expression given by Eq.~\eqref{eq:path_count} indeed converges to the simulated logical failure rate as $p\to 0$. We also observed that the path-counting expression (Eq.~\eqref{eq:path_count}) underestimates the logical failure rate when $p>1/L^2$. This regime of validity agrees with the analytical arguments in Appendix~\ref{app:path_count}. Thus we find that the regime of validity of the path counting approximation to the logical failure rate \textit{is much lower} than in the fully-post-selected case. This is presumably due to the entropic contributions of the much larger number of error configurations that can occur without post-selection.

\subsection{Below-threshold scaling}\label{subsec:subthreshold_nopostselect}

As the physical error rate $p$ increases beyond the narrow applicability of path-counting but remains below the critical threshold $p_{c}$ of the RBIM, the non-post-selected surface code enters the below-threshold regime. Analogous to the below-threshold regime of the fully post-selected surface code, we expect logical failure rates $\mathbb{P}_{\mathrm{fail}}$ in this regime to decay exponentially with code distance $L$. In the presence of quenched disorder, however, this decay is controlled not by the disorder-averaged domain wall free energy itself, but by the rare event tail of the distribution of domain wall free energy costs $\Delta F_{m}(p,L)$.

When quenched disorder is introduced, the capillary wave fluctuations described in Section.~\ref{subsec:below-threshold-postselect} are suppressed because the disorder introduces random pinning sites that cause the domain wall to become rougher and localized rather than thermally fluctuating \cite{PhysRevLett.54.2708}. The entropy of the interface becomes dominated by the energy cost of overcoming disorder pinning rather than the entropic wandering of the domain wall, effectively rescaling the surface tension to a \emph{smaller} value\footnote{More generally, in the generalized Ising model \cite{10.1063/1.1665530} of decoding, duality and GKS inequalities \cite{10.1063/1.1705219,10.1063/1.1664600} imply that the conditional MLD success probability is maximized in the clean (fully post-selected) case. Equivalently, in the convention used here, the corresponding domain wall free energy cost is at most as large as its clean model value. See Eq.~(18) and Appendix A of Ref.~\cite{PhysRevA.97.062320}.} and eliminating the $\log L$ entropic correction from capillary waves.

Rather than the deterministic domain wall free energy cost obtained in the fully post-selected model, the non-post-selected setting produces a distribution of domain wall free energy costs. At low temperatures, the scaling of this distribution is naturally motivated by directed polymer in a random environment (DPRE) theory \cite{PhysRevLett.54.2708,PhysRevLett.58.2087,PRXQuantum.4.010331}. The geometry relevant here, however, is not precisely the conventional point-to-point or point-to-line polymer geometry \cite{PRXQuantum.4.010331}: the domain wall is free to choose its position rather than having either endpoint fixed. We therefore use the DPRE picture only as a qualitative guide to the expected free energy distribution, and hence to the functional form of the logical failure rates observed numerically below.

In the DPRE picture, $\Delta F$ is most easily expressed as a function of a random variable $\chi$ which follows a Tracy-Widom law \cite{Tracy1994,Tracy1996}. The functional form of $\Delta F$ depends both on (inverse) temperature and the length of the domain wall, and is predicted to scale as
\begin{equation}
    \Delta F(L,\beta)\sim\sigma L - cL^{1/3}\chi + \gamma + \cdots,
\end{equation}
where $\sigma$ and $c$ are $\beta$-dependent constants, $\chi$ follows a Tracy-Widom law and $\cdots$ denotes subleading corrections. We treat the additional distribution $\gamma$ as a constant here, which can be justified by the fact that it is tightly distributed (see e.g. Eqs.~(1.5-1.7) of Ref.~\cite{Ferrari_2011}). With logical failure occurring whenever the domain wall free energy cost is negative, the leading asymptotics of $\mathbb{P}_{\mathrm{fail}}$ are set by
\begin{align}\label{eq:Pfail_DPRE}
    \mathbb{P}_{\mathrm{fail}}&=\mathbb{P}(\Delta F<0)\nonumber\\
    &=\mathbb{P}(\sigma L - cL^{1/3}\chi + \gamma<0)\nonumber\\
    &=\mathbb{P}\left(\chi > \frac{\sigma}{c}L^{2/3} + \frac{\gamma}{c}L^{-1/3}\right).
\end{align}
The right tail asymptotics of the Tracy-Widom distributions obey \cite{Dumaz_2013}
\begin{equation}\label{eq:TW_right_tail}
    \mathbb{P}(\chi > x) \sim \frac{c'}{x^{3/2}}e^{-c''x^{3/2}}.
\end{equation}
By substituting Eq.~\eqref{eq:TW_right_tail} into Eq.~\eqref{eq:Pfail_DPRE}, these asymptotics show that $\log(\mathbb{P}_{\mathrm{fail}})$ obtains a term linear in $L$. The subleading terms are logarithmic in $L$ and constant, respectively. In full, we have
\begin{align}
    &\log(\mathbb{P}_{\mathrm{fail}})=-c''\left(\frac{\sigma}{c}\right)^{3/2}L - \log L\nonumber\\
    &+ \left[\log c' - \frac{3}{2}\log\left(\frac{\sigma}{c}\right) - \frac{3c''\gamma\sqrt{\sigma}}{2c^{3/2}} \right] + \mathcal{O}(L^{-1}).
\end{align}
In other words, the leading asymptotic of $\mathbb{P}_{\mathrm{fail}}$ is an exponential decay in $L$, consistent with numerical observations in the literature \cite{PhysRevA.88.062308,Beverland_2019}. We expect both the finite-temperature (MLD) and zero-temperature (MWPMD) domain wall energy cost distributions to follow this form, as the universal aspects of the DPRE are the same in both settings \cite{PRXQuantum.4.010331,PhysRevLett.55.2923}. We leave a detailed investigation of this DPRE-related scaling to future studies.

The quenched disorder present in the RBIM prohibits a simple analytic evaluation of the true domain wall surface tension. However, by the arguments which predict the asymptotic scaling of $\log(\mathbb{P}_{\mathrm{fail}})$ above, we postulate that over the numerically accessible finite-size regime, the logical failure rate of a single logical degree of freedom takes the effective form
\begin{equation}\label{eq:sigma_eff}
    \mathbb{P}_{\mathrm{fail}}(p,L)\approx e^{-\beta[\sigma_{\mathrm{eff}}(p)L+\delta(p)]}.
\end{equation}
Here $\sigma_{\mathrm{eff}}(p)$ plays the role of an \emph{effective tension}, and $\delta(p)$ is a finite-size offset that absorbs subleading corrections over the range of lattice sizes considered. This effective tension $\sigma_{\mathrm{eff}}(p)$ is \emph{not} equal to the actual surface tension of the random-bond Ising model, but rather a function of the surface tension and the amplitude of DPRE fluctuations, which we define phenomenologically to allow equal comparisons between non-post-selected and post-selected surface codes.

This parametrization is intended as an effective finite-size description of the data rather than an asymptotically exact statement about the full disorder-averaged free energy. In particular, any logarithmic or other subleading corrections to $\log(\mathbb{P}_{\mathrm{fail}})$ are absorbed into $\delta(p)$ over the finite range of system sizes accessible in our numerics.

Although we cannot directly evaluate $\sigma_{\mathrm{eff}}(p)$ in our model, we determine its value in the limit as $p\to 0$ in the path-counting regime in Appendix~\ref{app:asymptotics}. The effective tension $\sigma_{\mathrm{eff}}(p)$ must still vanish at criticality, as with the surface tension of the clean Ising model \cite{PhysRev.65.117}. If we consider the behaviour precisely at threshold, Eq.~\eqref{eq:sigma_eff} implies $\mathbb{P}_{\mathrm{fail}}(p_{c},L)\approx e^{-\beta\delta(p_{c})}$. This finite-size offset is therefore expected to approximately determine the lattice size-independent failure rate at threshold. However, we note that the validity of these below-threshold approximations does not extend all the way to the threshold. 

\begin{figure}[t]
    \centering
    \includegraphics[width=\linewidth]{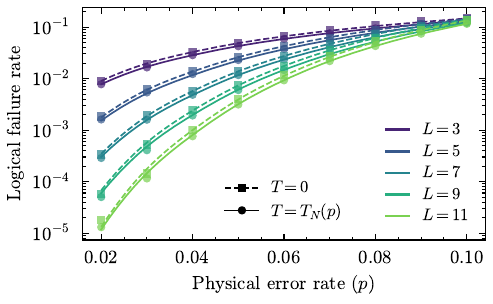}
    \caption{Comparison of the logical failure rate $\mathbb{P}_{\mathrm{fail}}$ for the non-post-selected planar code under bit-flip noise in the below-threshold regime, evaluated at zero temperature ($T=0$) and along the Nishimori line ($T=T_{\mathrm N}(p)$). For each dataset, the failure rate is fitted to the ansatz in Eq.~\eqref{eq:sigma_eff}, yielding the parameters $\sigma_{\mathrm{eff}}(p)$ and $\delta(p)$. The scatter points show the numerical results, while the fitted curves are shown as lines: dashed for $T=0$ and solid for $T=T_{\mathrm N}(p)$. The $T=0$ data are obtained using \emph{PyMatching}~\cite{pymatchingv2}, while the $T=T_{\mathrm N}(p)$ data are obtained using \emph{Planar}~\cite{PhysRevLett.134.190603}, both with $n=10^{6}$ samples.}
    \label{fig:curve_fit}
\end{figure}

\begin{figure}[t]
    \centering
    \includegraphics[width=\linewidth]{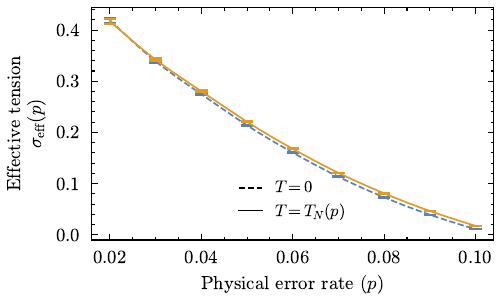}
    \caption{Plot of the effective tension $\sigma_{\mathrm{eff}}$ as a function of the physical error rate $p$ for the non-post-selected planar code in the below-threshold regime, comparing zero temperature ($T=0$) with the Nishimori line ($T=T_{\mathrm N}(p)$). The points are obtained from fits to the logical failure rate data, with error bars indicating one standard deviation, and the lines are interpolations between the fitted values: dashed for $T=0$ and solid for $T=T_{\mathrm N}(p)$.}
    \label{fig:sigmaeff}
\end{figure}

\begin{figure}[t]
    \centering
    \includegraphics[width=\linewidth]{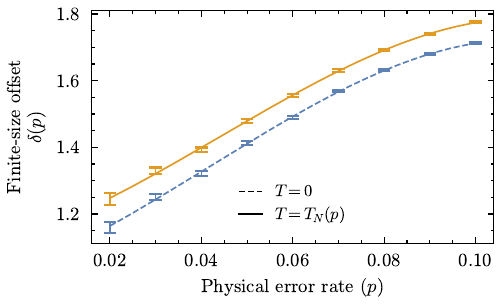}
    \caption{Plot of the finite-size offset $\delta(p)$ as a function of the physical error rate $p$ for the non-post-selected planar code in the below-threshold regime, comparing zero temperature ($T=0$) with the Nishimori line ($T=T_{\mathrm N}(p)$). The points are obtained from fits to the logical failure rate data, with error bars indicating one standard deviation, while the lines are guides to the eye generated by a weighted least-squares cubic polynomial fit: dashed for $T=0$ and solid for $T=T_{\mathrm N}(p)$.}
    \label{fig:deltap}
\end{figure}

We curve-fit and plot the logical failure rate of non-post-selected planar codes as a function of physical error rate and code distance in Fig.~\ref{fig:curve_fit} at both $T=0$ (through MWPMD) and $T=T_{N}(p)$ (through MLD). From this we extract the effective tension $\sigma_{\mathrm{eff}}(p)$, plotted in Fig.~\ref{fig:sigmaeff}, and the finite-size offset $\delta(p)$, plotted in Fig.~\ref{fig:deltap}. These fits show that the effective finite-size form in Eq.~\eqref{eq:sigma_eff} provides a good description of the below-threshold failure statistics of the planar code over the range of system sizes considered here at both $T=T_{N}(p)$ and $T=0$.

In Fig.~\ref{fig:sigmaeff}, the effective tension $\sigma_{\mathrm{eff}}(p)$ appears to be very similar at $T=0$ and $T=T_{N}(p)$, with any differences remaining small on the scale of the fitted uncertainties. By contrast, Fig.~\ref{fig:deltap}) shows that $\delta(p)$ at $T=0$ and $T=T_{N}(p)$ differ by an approximately constant offset across the range of $p$ studied here. Taken together, these results suggest that MWPMD and optimal MLD exhibit similar below-threshold scaling, with the dominant exponential suppression governed by nearly the same effective tension and the main discrepancy captured by a subleading finite-size offset. If this behaviour persists at larger $L$, then the resulting difference in logical failure rates should become negligible relative to the additional computational cost of optimal MLD.

By casting our results into the standard parametrization of Bravyi and Vargo \cite{PhysRevA.88.062308}, we give the decay rate $\alpha(p)$ a clear physical interpretation in terms of an effective tension governed by the free energy cost of rare logical domain walls. For our model, in the below-threshold regime of the surface code, the logical failure rate takes the form of Eq.~\eqref{eq:sigma_eff}. Our effective tension maps directly to the decay rate:
\begin{equation}\label{eq:decay_rate}
        \alpha(p)\vcentcolon=-\lim_{L\to\infty}\frac{1}{L}\log\left(\mathbb{P}_{\mathrm{fail}}(p,L)\right)=\beta \sigma_{\rm eff}(p).
\end{equation}

While the ansatz given by Eq.~\eqref{eq:sigma_eff} provides a good description of the logical failure rate deep within the below-threshold regime, it is not expected to hold as the physical error rate $p$ approaches the critical point $p_{c}$. To accurately describe the logical failure rates in this critical regime, we turn to the framework of finite-size scaling theory.

\subsection{Near-threshold}\label{subsec:nearthresh_nopostselect}

In the near-threshold regime, the behaviour of the logical failure rate $\mathbb{P}_{\mathrm{fail}}$ is governed by finite-size scaling. A consequence of introducing quenched disorder, characteristic of the random-bond Ising model relevant to non-post-selected surface codes, is that the free energy cost of a domain wall is no longer a single well-defined value for a given $p$ and $L$. In what follows, we study this distribution in two decoding settings: in the zero-temperature limit $T=0$, corresponding to MWPMD, where the relevant quantity is the domain wall ground state energy cost $\Delta E$, and along the Nishimori line $T=T_{N}(p)$, corresponding to optimal MLD, where the relevant quantity is the domain wall free energy cost $\Delta F$. In the quantum error correction literature, the former quantity has been termed the ``logical gap'', since it measures the difference in Hamming weight between decoder corrections leading to distinct logical outcomes \cite{gidney2023yokedsurfacecodes,PRXQuantum.5.010302,Smith_2024}\footnote{Under the statistical mechanical mapping, the domain wall ground state energy cost is directly proportional to the logical gap. Concretely, at zero temperature each violated bond contributes an energy cost of $2J$, so the ground state energy cost of inserting a nontrivial domain wall is $2\times(\text{logical gap})$. In this sense the logical gap is the combinatorial counterpart of the domain wall energy.}. At the critical point the mean of the distribution of domain wall energies $\Delta E$ is known to scale with $L$ as $\Delta E\propto L^\theta$ \cite{pandey2026} where $0<\theta<1$ is the stiffness exponent that has been studied in the spin glass literature \cite{PhysRevB.75.174415,Fisch2006,PhysRevB.97.064410,PhysRevLett.91.087201,N.Kawashima_1997,PhysRevB.70.134425}. Below, we analyze finite-size scaling results for both the full energy cost distributions and the logical error rates close to threshold. 

To simplify the initial presentation of our analysis, we primarily consider a planar code with open boundary conditions. This system encodes a single logical qubit $(k=1)$, in contrast to the two logical qubits of the toric code, allowing for a clearer exposition of the core concepts. The framework, however, extends straightforwardly to systems with multiple logical qubits, where it necessitates working with multi-dimensional probability distributions for the logical operator costs.

The relevant domain wall costs in the RBIM can take either sign due to the random couplings in the model. We denote this quantity by $\Delta \mathcal{C}$, where $\Delta \mathcal{C}=\Delta E$ at $T=0$, and $\Delta \mathcal{C}=\Delta F$ along the Nishimori line ($T=T_{N}(p)$). In the quantum error correction context negative domain wall energy corresponds to decoder failure. We will define the probability density for $\Delta \mathcal{C}$ to be $\mathbb{P}(\Delta \mathcal{C};p,L)$. As a result the logical failure probability is just the total probability for negative $\Delta \mathcal{C}$ and can be determined if $\mathbb{P}(\Delta \mathcal{C};p,L)$ is known. We will assume as in Eq.~\eqref{eq:genscaling} that the logical failure probability is given as a function of the scaling variable $x=(p-p_{c})L^{{1/\nu}}$ where $\nu$ is the scaling exponent of the RBIM. Thus we have
\begin{align}\label{eq:errordist}
    \mathbb{P}_{\mathrm{fail}}(p,L) & = \int_{-\infty}^{0}d\Delta \mathcal{C}\ \mathbb{P}(\Delta \mathcal{C};p,L)
    \nonumber \\
    &=f((p-p_{c})L^{\frac{1}{\nu}}),
\end{align}
where $f$ is the scaling function. The general scaling function $f(x)$, which represents $\mathbb{P}_{\mathrm{fail}}(x)$, is expected to satisfy asymptotic limits $f(x)\to0$ as $x\to-\infty$ (below-threshold regime, $p<p_{c}$) and $f(x)\to (K-1)/K$ as $x\to\infty$ (above-threshold regime, $p>p_{c}$) \cite{WangChenyang2003Ctia}, where $K$ is the dimension of the logical subalgebra. For the planar code with bit-flip noise, which encodes a single logical qubit ($k=1$), the upper limit is $(2-1)/2=1/2$ (see Appendix \ref{app:supplementary_scaling}, Sec.~\ref{app:asymptotics_gx}).

Eq.~\eqref{eq:errordist} is highly constraining for the energy cost distributions at threshold. At the critical point $p=p_{c}$, the diverging correlation length renders the model scale invariant, and this is reflected in a domain wall energy distribution that becomes, in a coarse-grained sense, independent of the lattice size. In particular, the condition that $\mathbb{P}_{\mathrm{fail}}$ remains constant for all finite $L$ reflects this scale invariance. 

\begin{figure}
    \centering
    \includegraphics[width=\linewidth]{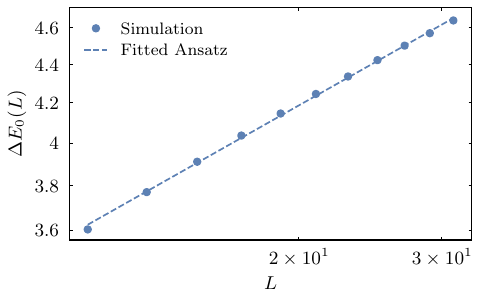}
    \caption{Log-log plot of the mean ground state energy cost, $\Delta E_{0}(L)$, between the two logical sectors of the planar code at criticality $p=p_{c}$ as a function of lattice size $L$. The scatter points represent the numerically simulated averages, and the solid line is a linear regression fit based on the ansatz $\Delta E_{0}(L)=AL^{\theta}$. The fit yields a stiffness exponent of $\theta=0.242$.}
    \label{fig:stiffness}
\end{figure}

At $T=0$, we verify numerically that the appropriately rescaled $\mathbb{P}(\Delta E;p,L)$ distributions converge to a universal envelope as $L\to\infty$ (see Appendix \ref{app:supplementary_scaling}, Secs.~\ref{app:scaling_form}-\ref{app:convergence_envelope}). Fitting $\langle \Delta E\rangle = A L^{\theta}$ using $n=10^6$ samples per point and $p_c=0.103$~\cite{WangChenyang2003Ctia} gives $\theta = 0.242$ (see Fig.~\ref{fig:stiffness}). One complexity arises from the fact that by virtue of the quantized coupling constants, the distribution $\mathbb{P}(\Delta E;p,L)$ is discrete for all $p\in [0,1], L\in \mathbb{N}^+$. The effect of this can be seen in Fig.~\ref{fig:rescaled_prob}. Consequently in the Appendix we are careful to define a suitably coarse-grained notion of convergence of these distributions.

At $T=0$, rescaling the distributions over $\Delta E$ at $p_{c}$ by $L^\theta$ collapses the data for all $L$ onto a single curve, well described (in the region relevant to the $\mathbb{P}_{\mathrm{fail}}$ integral) by a Gaussian with mean $\mu=2.047$ and standard deviation $\sigma=2.010$ (see Fig.~\ref{fig:rescaled_prob} and Appendix \ref{app:supplementary_scaling}, Sec.~\ref{app:scaling_form}). This motivates a Gaussian approximation for $\mathbb{P}(\Delta E;p,L)$ in what follows.

\begin{figure}[t]
    \centering
    \includegraphics[width=\linewidth]{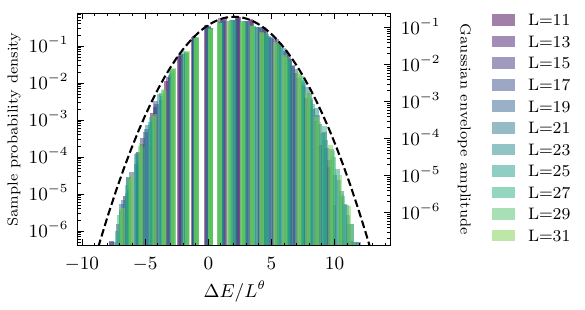}
    \caption{Scaled probability distributions of the domain wall ground state energy costs $\Delta E(L)/L^{\theta}$ for the planar code at criticality $(p=p_{c}\approx 0.103)$. Each histogram (coloured) corresponds to a different lattice size $L$, and the overlaid dashed curve is a single Gaussian whose mean and standard deviation are taken from the pooled data of $\Delta E/L^{\theta}$ ($\mu=2.047$, $\sigma=2.010$). The Gaussian aims to approximate the proposed continuous envelope onto which all distributions collapse as $L\to\infty$.}
    \label{fig:rescaled_prob}
\end{figure}

The ground state energy cost distribution for the 2D $\pm J$ random-bond Ising model with open boundary conditions (i.e., the planar code) is known to exhibit non-Gaussian tails \cite{PhysRevB.75.174415}. This non-Gaussianity can be clearly seen in the tails of Fig.~\ref{fig:rescaled_prob}. However, for the purpose of determining the logical failure rate $\mathbb{P}_{\mathrm{fail}}$ via integration of this distribution, and that of the ($T=T_{N}(p)$) free energy cost distribution, we find that its behaviour, particularly in the region relevant to the integral, can be approximated to leading order by a transformed Gaussian. While this approximation neglects the true non-Gaussian nature of the tails, which can introduce non-universal, $L$-dependent corrections, it captures the dominant scaling behaviour. 

\begin{figure}[t]
    \centering
    \includegraphics[width=\linewidth]{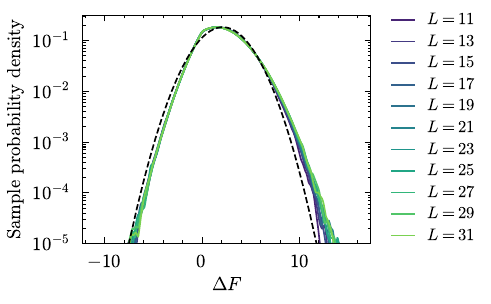}
    \caption{Probability distributions of the domain wall free energy cost, $\Delta F(L)$, for the two logical sectors of the planar code at criticality $(p = p_{c}\approx0.1092212)$, evaluated on the Nishimori line $T = T_{N}(p)$. Each coloured curve corresponds to a different lattice size $L$. Probability density functions at each $L$ are estimated using a non-parametric Gaussian kernel density estimator (KDE). The dashed black curve shows a single Gaussian with mean and standard deviation taken from the pooled data of $\Delta F(L)$ ($\mu = 2.115$ and $\sigma = 2.167$). Without any additional rescaling by $L^{\theta}$ as in Fig.~\ref{fig:rescaled_prob}, the distributions exhibit an approximately scale invariant form, consistent with a stiffness exponent of $\theta=0$.}
    \label{fig:rescaled_prob_optimal}
\end{figure}

Conversely, at $T=T_{N}(p)$, the distribution of $\Delta F$ at $p_{c}$ is manifestly scale invariant without any rescaling. In other words, at this point, the stiffness exponent $\theta=0$ exactly. We show this distribution in Fig.~\ref{fig:rescaled_prob_optimal}, using $p=p_{c}\simeq0.1092212$ \cite{wan2025}. A Gaussian distribution is overlaid on the estimated probability distributions, with mean and standard deviation taken from the pooled data ($\mu=2.115$ and $\sigma=2.167$). The fit of this Gaussian is not as good as at $T=0$, due to the known asymmetry of the distribution of $\Delta F$~\cite{wan2025}. This distribution is well studied, and includes a non-zero skew and a nonanalytic kink at $\Delta F=0$. This kink arises from the Nishimori symmetry constraint~\cite{chen2025scalableaccuracygainspostselection}
\[
\mathbb{P}(\Delta F)=\mathbb{P}(-\Delta F)e^{\beta\Delta F}
\]
enforcing the left and right derivatives of $\mathbb{P}(\Delta F)$ to sum to the probability at the origin~\cite{wan2025}
\[
\partial_{-}(0)+\partial_{+}(0)=\mathbb{P}(0).
\]

\begin{figure*}[t]
    \centering
    \subfloat[\label{fig:data_collapse_plot_mu_sigma}]
    {\includegraphics[width=0.5\textwidth]{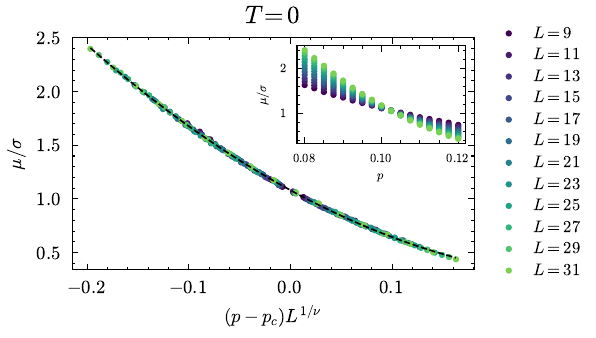}}
    \hfill
    \subfloat[\label{fig:data_collapse_plot_mu_sigma_optimal}]
    {\includegraphics[width=0.5\textwidth]{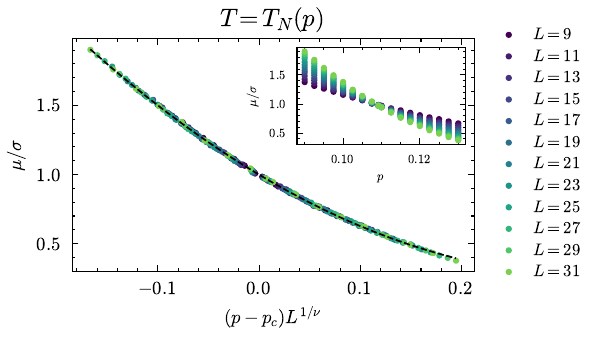}}

    \caption{
    Data collapse plots of the ratio of mean to standard deviation ($\mu/\sigma$) of domain wall costs as a function of the scaling variable
    $x=(p-p_c)L^{1/\nu}$.
    (a) Zero-temperature domain wall ground state energy cost distribution, obtained by MWPMD.
    The dashed black line is the best fit to the second-order polynomial ansatz $g(x)=B_2x^2+B_1x+B_0$.
    (b) Finite-temperature domain wall free energy cost distribution on the Nishimori line, obtained by optimal MLD. The dashed black line is the best fit to the same second-order polynomial ansatz. In both cases, the collapse of data from different system sizes onto a single curve supports the finite-size scaling hypothesis near the respective threshold. Note the different scale of the y-axes.}
    \label{fig:data_collapse_combined}
\end{figure*}

For the purposes of our analysis of $\mathbb{P}_{\mathrm{fail}}$, we approximate the domain wall energy cost distributions $\mathbb{P}(\Delta \mathcal{C};p,L)$ by a Gaussian with mean $\mu(p,L)$ and variance  $\sigma^2(p,L)$ for both $T=0$ and $T=T_{N}(p)$ in what follows for consistent analyses. The failure probability is obtained by integrating the negative tail of this Gaussian distribution:
\begin{align}\label{eq:Pfail_erf_general}
\mathbb{P}_{\mathrm{fail}}(p,L)\approx \frac{1}{2} \left[ 1 - \mathrm{erf}\left(\frac{\mu(p,L)}{\sigma(p,L)\sqrt{2}}\right) \right]. 
\end{align}
Given the assumed scaling behaviour of the logical failure probability, we expect the ratio $\mu(p,L)/\sigma(p,L)$ to be a function of the scaling variable $x=(p-p_{c})L^{1/\nu}$ near-threshold. Let us define
\begin{equation}\label{eq:g(x)}
    g(x)=\frac{\mu(p,L)}{\sigma(p,L)}.
\end{equation}
We analyze the energy cost distributions found in simulations close to the known threshold at $p=p_c\simeq0.103$ for $T=0$ \cite{WangChenyang2003Ctia} and at $p=p_{c}\simeq 0.1092212$ for $T=T_{N}(p)$ \cite{wan2025} for a range of system size $L$ to find the mean and variance of the distribution. We plot the ratio $g(x)$ in Figs.~\ref{fig:data_collapse_plot_mu_sigma} and \ref{fig:data_collapse_plot_mu_sigma_optimal} and observe very good data collapse, indicating the quality of our scaling ansatz. We curve-fit a second-order polynomial to the ratio $g(x)$, using the ansatz $g(x) = B_2 x^2 + B_1 x + B_0$. For $T=0$, the best-fit parameters extracted from this procedure are: $p_c = 0.1014 \pm 0.0001$, $\nu = 1.60 \pm 0.02$, $B_2 = 9.569 \pm 0.3$, $B_1 = -5.399 \pm 0.08$, and $B_0 = 1.091 \pm 0.004$. For $T=T_{N}(p)$, the best-fit parameters extracted from this procedure are: $p_c = 0.1085 \pm 0.00006$, $\nu=1.560 \pm 0.008$, $B_2 = 6.456 \pm 0.1$, $B_1 = -4.363 \pm 0.04$, and $B_0 = 0.9988 \pm 0.002$.

\begin{figure*}[t]
    \centering
    \subfloat[\label{fig:curvefit}]
    {\includegraphics[width=0.5\textwidth]{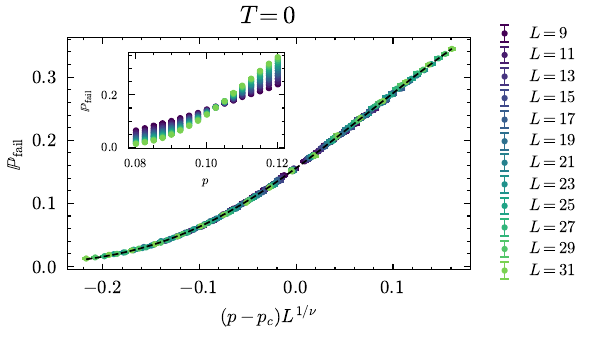}}
    \hfill
    \subfloat[\label{fig:curvefit_optimal}]
    {\includegraphics[width=0.5\textwidth]{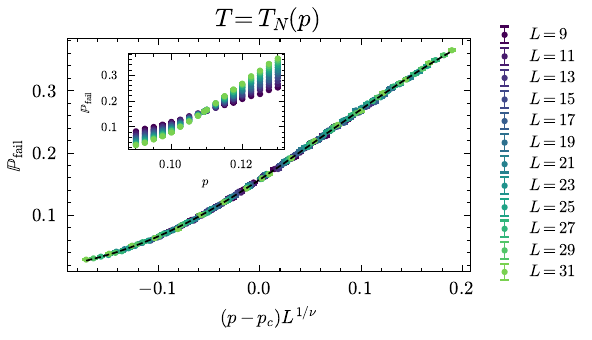}}

    \caption{
    Logical failure rate $\mathbb{P}_{\mathrm{fail}}$ for the non-post-selected planar code under bit-flip noise, plotted as a function of the scaling variable $x=(p-p_{c})L^{1/\nu}$ to demonstrate data collapse.
    (a) Zero temperature ($T=0$), corresponding to MWPMD.
    (b) Along the Nishimori line, $T=T_{\mathrm N}(p)$, near the Nishimori multicritical point, corresponding to optimal MLD.
    In both panels, the dashed black curve shows the best fit to the finite-size scaling ansatz of Eq.~\eqref{eq:FSS_general_erf_ansatz}, with quadratic scaling function $G(x)$, treating the parameters $A_{2},A_{1},A_{0}$, the critical error rate $p_{c}$, and the critical exponent $\nu$ as free fit parameters.
    The scatter points represent numerical simulation results obtained using \emph{PyMatching}~\cite{pymatchingv2} with $n=10^{5}$ samples in (a), and \emph{Planar}~\cite{PhysRevLett.134.190603} with $n=10^{5}$ samples in (b).}
    \label{fig:curvefit_combined}
\end{figure*}

To construct a practical ansatz for fitting numerical data for the logical failure rate, particularly for capturing the behaviour around the threshold, we use a phenomenological function inspired by the error function structure of Eq.~\eqref{eq:Pfail_erf_general}. The general form we consider is:
\begin{equation}\label{eq:FSS_general_erf_ansatz}
f_{\mathrm{erf}}(x) = \frac{1}{2}\biggl(1-\mathrm{erf}(G(x)/\sqrt{2})\biggr),
\end{equation}
where $G(x)$ is a polynomial in the scaling variable ${x=(p-p_c)L^{1/\nu}}$, and $p_c$ and $\nu$ are also fitting parameters. The order of the polynomial $G(x)$ determines the total number of free parameters in this ansatz.
Specifically, we investigate two cases for $G(x)$:
\begin{enumerate}
    \item \textbf{Quadratic $G(x)$}: $G_{\mathrm{quad}}(x) = A_2 x^2 + A_1 x + A_0$. This results in a 5-parameter model for $f_{\mathrm{erf}}(x)$ (namely $A_2, A_1, A_0, p_c, \nu$). This is the primary form used for the data collapse shown in Figs.~\ref{fig:curvefit} and \ref{fig:curvefit_optimal}.
    \item \textbf{Linear $G(x)$}: $G_{\mathrm{lin}}(x) = A_1 x + A_0$. This results in a 4-parameter model for $f_{\mathrm{erf}}(x)$ (namely $A_1, A_0, p_c, \nu$).
\end{enumerate}
We plot in Figs.~\ref{fig:curvefit} and \ref{fig:curvefit_optimal} a data collapse curve using the quadratic $G(x)$ ansatz (5 free parameters), rescaling by the argument $x=(p-p_{c})L^{\frac{1}{\nu}}$. At $T=0$, our fit returned values $A_2=5.509\pm 0.2$, $A_1=-4.613\pm 0.06$, and $A_0=1.041\pm0.002$. The fit also refined the critical parameters to $p_{c}=0.1028\pm0.0001$ and $\nu=1.517\pm0.01$, which are used to construct $x$ in the plot. At $T=T_{N}(p)$, our fit returned values $A_2=5.270 \pm 0.1$, $A_1=-4.466 \pm 0.04$, and $A_0 = 1.006 \pm 0.001$. The fit also returned the critical parameters as $p_{c}=0.1090 \pm 0.00004$ and $\nu=1.562 \pm 0.008$.

To quantify the critical failure rate, we compare three independent methods that map either the domain wall energy cost distribution or the logical failure data to $\mathbb{P}_{\mathrm{fail}}(p_c)$. We first investigate the critical failure rate at $T=0$ using MWPMD. \textbf{(1)} Fixing $p_c=0.103$ and fitting Gaussians to the critical distributions across $L$ gives $g(0)=\mu/\sigma=1.018 \Rightarrow \mathbb{P}_{\mathrm{fail}}(p_c)=0.154$. \textbf{(2)} Performing a data collapse on $g(x)=\mu/\sigma$ with $(p_c,\nu)$ free yields $p_c=0.1014\pm0.0001$ and an intercept $B_0=1.091\pm0.004 \Rightarrow \mathbb{P}_{\mathrm{fail}}(p_c)=0.1376\pm0.0008$. \textbf{(3)} Fitting the logical failure rate directly with the (quadratic) error-function ansatz gives $p_c=0.1028\pm0.0001$ and $A_0=1.041\pm0.002 \Rightarrow \mathbb{P}_{\mathrm{fail}}(p_c)=0.1489\pm0.0004$. The predictions of the critical failure rate do not agree within uncertainty. However, overall, all three methods place $\mathbb{P}_{\mathrm{fail}}(p_c)$ in the narrow range of $0.135$--$0.155$; we thus rely on the direct failure rate fit for precise threshold estimates and use the distribution-based collapse as a complementary cross-check of the scaling form.

We next investigate the critical failure rate at $T=T_{N}(p)$ using optimal MLD. \textbf{(1)} Fixing $p_c=0.1092212$ and fitting Gaussians to the critical distributions across $L$ gives $g(0)=\mu/\sigma=0.9760 \Rightarrow \mathbb{P}_{\mathrm{fail}}(p_c)=0.1645$. \textbf{(2)} Performing a data collapse on $g(x)=\mu/\sigma$ with $(p_c,\nu)$ free yields $p_c = 0.1085 \pm 0.00006$ and an intercept $B_0 = 0.9988 \pm 0.002 \Rightarrow \mathbb{P}_{\mathrm{fail}}(p_c)=0.1589 \pm 0.0005$. \textbf{(3)} Fitting the logical failure rate directly with the (quadratic) error-function ansatz gives $p_{c}=0.1090 \pm 0.00004$ and $A_0 = 1.006 \pm 0.001 \Rightarrow \mathbb{P}_{\mathrm{fail}}(p_c)=0.1572 \pm 0.0002$. Again, the predictions from these three methods do not agree within uncertainty. However, all three methods place $\mathbb{P}_{\mathrm{fail}}(p_c)$ in the narrow range of $0.155$--$0.165$.

We compared the quadratic error function ansatz (Eq.~\eqref{eq:FSS_general_erf_ansatz}) with several alternatives, including a linear error function form, and two commonly used ans\"{a}tze from literature: a simple polynomial model, and a polynomial with finite-size corrections (see Appendix~\ref{app:performance_comparison} for details). Across all three comparison metrics -- the residual sum of squares (RSS) and the Akaike and Bayesian information criteria (AIC and BIC) -- the quadratic error function model consistently gave the best fit to our numerical data for both the $T=0$ and $T=T_{N}(p)$ settings (see Tables~\ref{tab:ansatz_compare_all_p_revised} and \ref{tab:ansatz_compare_all_p_revised_optimal} in Appendix~\ref{app:performance_comparison}). This strong performance, despite using fewer parameters than some alternatives, highlights that the error function structure, motivated by the Gaussian approximation to the energy-cost distribution, more accurately captured the true scaling behaviour of $\mathbb{P}_{\mathrm{fail}}$ in the near-threshold regime. 

We believe that the improved performance of this ansatz results from the fact that our chosen scaling function ansatz fits the failure probability accurately throughout the region where the logical error probability changes rapidly with physical error probability. This region carries the most information about the location of the threshold even when the relevant range of physical error probability shifts as the size of the system increases. The usual approach, such as in Ref.~\cite{WangChenyang2003Ctia}, Taylor expands the scaling function close to the point where the logical failure curves cross. As we have seen, this can be quite far from the point where the derivative of the failure probability is maximum. More dramatic examples of this include recent work on dynamical bias arrangement~\cite{bombin2023} (see Fig.~8) and on the sustainable threshold in single-shot error correction~\cite{cjb4-l57n} (see Fig.~S2). In these cases the crossing point occurs where $\mathbb{P}_{\mathrm{fail}}\simeq 1$, making locating the threshold by means of the crossing less efficient. In general we believe that scaling ans\"{a}tze that fit the data accurately in the region where the derivative of the failure probability is largest will outperform those that look for the crossing. There are many possible approaches, an example is the ansatz described in section~S3.C of Ref.~\cite{cjb4-l57n}.

Finally, we do not include a dedicated above-threshold subsection for non-post-selected surface codes. Quenched disorder does not destroy the Kramers-Wannier duality; rather, it replaces the clean $p\leftrightarrow p^{*}$ map of the clean model with a generalized duality that no longer yields a simple relation between the RBIM at $p$ and any dual point at $p^{*}$ (see, e.g., Refs.~\cite{Song_2024,10.1063/1.1665530}). Consequently, we cannot simply bootstrap an above-threshold ansatz (e.g., $\sigma',\delta'$) from the below-threshold quantities. Second, for $p\gg p_{c}$, the logical failure rate rapidly saturates toward $(K-1)/K$ with increasing $L$, making differences between code distances exponentially hard to resolve and of limited diagnostic value; only in the narrow near-threshold window governed by finite-size scaling does $\mathbb{P}_{\mathrm{fail}}$ carry useful `distance-to-threshold' information.

\section{Extension to circuit-level noise}\label{sec:circuit-level}

In this section, we investigate how the insights developed above can be extended beyond code capacity noise, by considering a rotated surface code memory experiment under uniform depolarizing circuit-level noise with $d$ rounds of syndrome extraction. In the presence of noisy syndrome measurements, the decoding problem is naturally associated with a three-dimensional statistical mechanical model. For independent bit-flip data errors and measurement errors this model is most easily formulated as the random plaquette gauge model (RPGM) \cite{10.1063/1.1499754,WangChenyang2003Ctia}. For depolarizing noise, the $Y$-error correlations couple the $X$ and $Z$ spacetime volumes, leading to the random coupled-plaquette gauge model (RCPGM) proposed in Ref.~\cite{rispler2024}. Circuit-level depolarizing noise does not map exactly to a simple uniform RCPGM, because microscopic circuit faults propagate into schedule-dependent correlated events. Nevertheless, the RCPGM provides a close statistical mechanical analogue for the depolarizing circuit-level setting, and motivates interpreting the circuit-dependent logical failure rates in terms of effective domain wall energy costs.

To investigate how well our ans\"atze can capture the behaviour of logical failure rates under circuit-level noise, we compare three syndrome extraction circuits: a standard hook error-avoiding $N/Z$ measurement schedule~\cite{PhysRevA.90.062320,Acharya2023,Geher2024errorcorrected}, the recently proposed diagonal schedule of Ref. \cite{kishony2026}, and a deliberately suboptimal hook-prone schedule in which correlated faults are aligned with the logical operators \cite{hirai2026}. The diagonal measurement schedule is designed to preserve the full circuit distance while maintaining a globally uniform measurement schedule. By contrast, in the hook-prone schedule the path-counting picture predicts a reduced effective distance: in the path-counting regime, the dominant fault patterns have support scaling like $d/2$ rather than $d$ \cite{kishony2026}. In the below-threshold regime, however, degeneracy and finite-$p$ effects soften this simple factor-of-two picture, so the effective suppression rate must be extracted numerically.

All circuit-level simulations were performed using \emph{Stim} \cite{gidney2021stim}, and decoding was carried out using \emph{PyMatching} \cite{pymatchingv2}. Motivated by the below-threshold form developed in Sec.~\ref{subsec:subthreshold_nopostselect}, we fit the logical failure rate for each schedule $s$ to
\begin{equation}\label{eq:circuit_ansatz}
    \mathbb{P}_{\mathrm{fail}}^{(s)}(p,d)
\approx A_s(p)\exp[-\alpha_s(p)d],
\end{equation}
where $\alpha_s(p)$ is a schedule-dependent effective decay rate, and $A_s(p)$ is a prefactor absorbing subleading finite-size and circuit-dependent effects.

Although the RCPGM has Nishimori conditions, these conditions specify multiple dimensionless coupling constants rather than a unique scalar inverse temperature. Moreover, the circuit-level noise model considered here contains schedule-dependent correlated fault mechanisms, so there is no unique Nishimori inverse temperature $\beta_{\mathrm N}(p)$ by which to normalize $\alpha_{s}(p)$ to obtain an effective tension. We therefore report $\alpha_{s}(p)$ itself as the primary phenomenological quantity.

\begin{figure*}[t]
    \centering
    \subfloat[]
    {\includegraphics[width=0.33\textwidth]{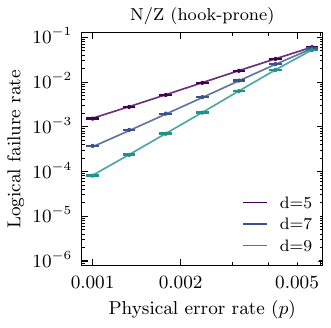}}
    \hfill
    \subfloat[]
    {\includegraphics[width=0.33\textwidth]{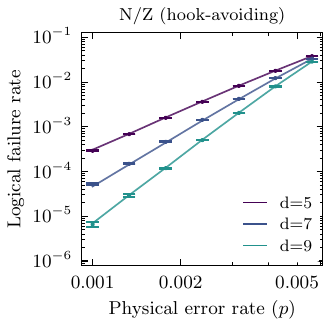}}
    \hfill
    \subfloat[]
    {\includegraphics[width=0.33\textwidth]{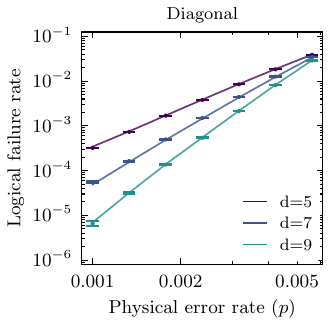}}
    \caption{Logical failure rate of a rotated surface code memory experiment under uniform depolarizing circuit-level noise. The code distance is $d$, and each experiment uses $d$ rounds of syndrome extraction, with $n=10^{7}$ shots per data point. Results are shown for three measurement schedules: (a) N/Z (hook-prone), (b) N/Z (hook-avoiding), and (c) diagonal. Scatter points show numerical simulation data and solid curves show fits to the phenomenological below-threshold ansatz.
    }
    \label{fig:circuit_level_Pfail}
\end{figure*}

Fig.~\ref{fig:circuit_level_Pfail} shows the logical failure rate as a function of the physical circuit depolarizing probability $p$, together with fits to the ansatz in Eq.~\eqref{eq:circuit_ansatz}. Across the below-threshold regime, the ansatz captures the numerical data well. The main effect of changing the syndrome extraction schedule is to change the decay rate: the hook-prone $N/Z$ schedule shows systematically weaker exponential suppression, while the hook-avoiding $N/Z$ and diagonal schedules exhibit larger and mutually consistent decay rates within uncertainty. This is consistent with the expectation that the hook-prone circuit lowers the effective cost of logical fault configurations, while the distance-preserving schedules restore the full circuit distance in the low-$p$ limit.

\begin{figure}[t]
    \centering
    \includegraphics[width=\linewidth]{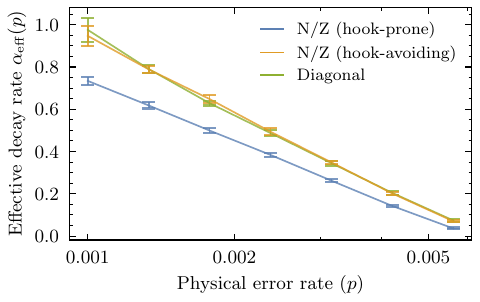}
    \caption{Fitted circuit-dependent decay rates $\alpha_{s}(p)$ extracted from the logical failure data in Fig.~\ref{fig:circuit_level_Pfail}. The decay rate decreases toward zero as $p$ approaches threshold, while at lower $p$ the three schedules separate according to circuit quality. The hook-prone schedule has the smallest decay rate, reflecting its reduced effective distance in the low-$p$ limit.}
    \label{fig:circuit_level_decay}
\end{figure}

To make this more explicit, in Fig.~\ref{fig:circuit_level_decay} we plot the fitted decay rates $\alpha_{s}(p)$. As $p$ increases toward threshold, the three curves move toward zero, consistent with the loss of exponential suppression near criticality being controlled mainly by the bulk transition. At smaller $p$, however, the curves separate clearly according to circuit quality. In particular, the standard $N/Z$ and diagonal schedules retain an equivalent (up to uncertainty) decay rate, whereas the hook-prone schedule has a smaller decay rate throughout the below-threshold regime. However, the hook-prone decay rate is larger than the $1/2$ reduction predicted by the path-counting picture.

\begin{figure*}[t]
    \centering
    \subfloat[]
    {\includegraphics[width=0.33\textwidth]{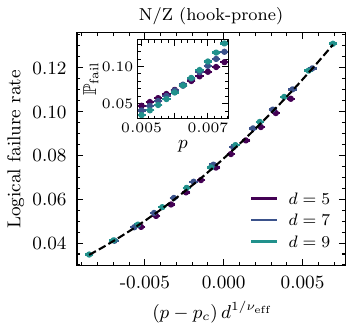}}
    \hfill
    \subfloat[]
    {\includegraphics[width=0.33\textwidth]{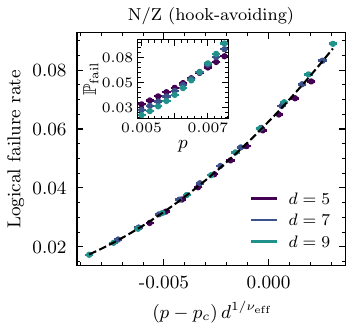}}
    \hfill
    \subfloat[]
    {\includegraphics[width=0.33\textwidth]{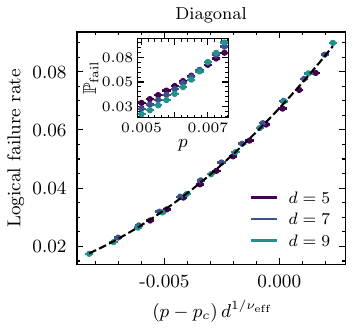}}
    \caption{Finite-size scaling collapse of the logical failure rate for the rotated surface code memory experiment under uniform depolarizing circuit-level noise. Results are shown for three syndrome extraction schedules: (a) N/Z (hook-prone), (b) N/Z (hook-avoiding), and (c) diagonal. For each schedule, the threshold $p_c^{(s)}$ and effective crossover exponent $\nu_{\mathrm{eff}}^{(s)}$ are fitted independently, and the scaling function is approximated locally by a second-order polynomial in the scaling parameter $x_{s}=(p-p_{c}^{(s)})d^{1/\nu_{\mathrm{eff}}^{(s)}}$.}
    \label{fig:circuit_level_data_collapse}
\end{figure*}

We next examine the near-threshold regime. For each schedule, we fit the logical failure rate to the finite-size scaling form
\[
\mathbb{P}_{\mathrm{fail}}^{(s)}(p,d)
\simeq
f_s\!\left((p-p_c^{(s)})d^{1/\nu_{\mathrm{eff}}^{(s)}}\right),
\]
where the fitted threshold $p_c^{(s)}$, effective crossover exponent $\nu_{\mathrm{eff}}^{(s)}$, and scaling function $f_s$ are all allowed to depend on the syndrome extraction schedule. The resulting collapses, shown in Fig.~\ref{fig:circuit_level_data_collapse}, are remarkably good for each schedule.

The fitted threshold and crossover exponent for each schedule are summarized in Tab.~\ref{tab:circuit_level_fss}. Both the threshold $p_{c}^{(s)}$ and the effective crossover exponent $\nu_{\mathrm{eff}}^{(s)}$ are maximized for the diagonal measurement schedule, minimized for the hook-prone schedule and lie in between for the hook-avoiding schedule. The fitted thresholds differ between schedules, as expected from the schedule-dependent propagation of microscopic circuit faults.

\begin{table}[t]
    \begin{tabular}{lcc}
        \hline
        \hline
        Schedule & $p_c^{(s)}$ & $\nu_{\mathrm{eff}}^{(s)}$ \\
        \hline
        N/Z (hook-prone) & $0.00621 \pm 0.00002$ & $1.20 \pm 0.03$ \\
        N/Z (hook-avoiding) & $0.00668 \pm 0.00002$ & $1.30 \pm 0.02$\\
        Diagonal & $0.00682 \pm 0.00003$ & $1.39 \pm 0.04$ \\
        \hline
        \hline
    \end{tabular}
    \caption{Fitted near-threshold parameters for the rotated surface code memory experiment under uniform depolarizing circuit-level noise.}
    \label{tab:circuit_level_fss}
\end{table}

The interpretation of the exponents $\nu_{\mathrm{eff}}^{(s)}$ requires care. In the RPGM and RCPGM literature, the underlying phase transition is often diagnosed using gauge-invariant observables such as Wilson loops \cite{WangChenyang2003Ctia,Andrist_2011} or Polyakov lines \cite{rispler2024}. Moreover, the RCPGM transition has been argued to be first order, in which case the correlation length remains finite and the usual interpretation of $\nu$ as a divergent correlation length critical exponent is not applicable \cite{Andrist_2011,rispler2024}. We therefore denote the fitted exponent by $\nu_{\mathrm{eff}}$ and interpret it only as a finite-size crossover exponent governing the logical failure probability. In this sense, the collapse should not be read as evidence for a conventional continuous critical point. Rather, it shows that the nonlocal QEC observable $\mathbb{P}_{\mathrm{fail}}$ exhibits a simple and robust finite-size scaling, even when the underlying statistical mechanical transition may be first order.

\section{Discussion and Outlook}\label{sec:discuss}

In this work, we have leveraged the exact solvability of the 2D Ising model to provide a comprehensive analytical treatment of the fully post-selected toric code, yielding closed-form expressions for its logical failure rate across the entire domain of code distances and physical error rates. This analysis established the existence and characteristics of four distinct operational regimes: path-counting, below-threshold (ordered), near-threshold (critical), and above-threshold (disordered), see Fig.~\ref{fig:combined_figure}. We derived approximations of the functional behaviour of the logical failure rates within each regime from a physical perspective. Specifically, we proposed a combinatorial expression counting the minimal uncorrectable errors in the path-counting regime. In the below-threshold regime, we incorporated capillary wave theory to propose an expression for the failure rates which depend on the surface tension of domain walls. Near criticality, we showed how finite-size scaling manifests in the logical failure rate through the domination of low-momentum modes of the partition functions with different boundary conditions. Finally, we showed how a Kramers-Wannier duality exactly mapped the failure rate in the below-threshold regime to a dual probability in the above-threshold regime.

These physical considerations then motivated analyses without post-selection, that is, on a standard surface code. First, we showed that path-counting expressions are accurate for non-post-selected toric codes, however, the accuracy of such approximations only hold for much lower physical error rates than suggested in prior numerical studies \cite{Watson_2014}. We next considered the physical behaviour of domain wall fluctuations within the below-threshold (ordered) regime to propose an ansatz for the failure rates. Our physically-motivated ansatz effectively captured the functional behaviour of failure rates. While recovering forms similar to previous ans\"{a}tze proposed in the literature (e.g., Ref.~\cite{Beverland_2019}), our approach provides a clearer physical interpretation of the constituent parameters in terms of effective tension-driven free energy costs. Finally, we investigated the near-threshold behaviour of failure rates by proposing a new notion of data collapse for the distribution of ground state energy costs of a random-bond Ising model. The universal envelope of this distribution was used to propose a new ansatz for the finite-size scaling function of the logical failure rate of surface codes near threshold. We compared this new error function-based ansatz against an exponential ansatz proposed in literature, obtaining an order-of-magnitude improvement in the fitting accuracy with equivalent numbers of free parameters.

The consistent structure of these four operational regimes across both the exactly solvable post-selected model and the numerically-investigated non-post-selected model suggests a broadly applicable framework for understanding topological QEC. We now delineate the general conditions under which these four regimes are expected to be a generic feature of such codes.

\subsection{Physical interpretation}

Our quantitative analysis of the non-post-selected toric code via the ansatz $\mathbb{P}_{\mathrm{fail}}(p,L)\approx e^{-\beta[\sigma_{\mathrm{eff}}(p)L+\delta(p)]}$ (Eq.~\eqref{eq:sigma_eff}) provides insights into the impact of quenched disorder. The extracted effective tension $\sigma_{\mathrm{eff}}(p)$ at both $T=0$ and $T=T_{N}(p)$ (Fig.~\ref{fig:sigmaeff}), exhibits a monotonically decreasing trend with $p$, vanishing as $p\to p_{c}$ as expected. This behaviour quantifies how the energy cost of creating a system-spanning domain wall is systematically reduced by increasing disorder, yet remains substantial enough to ensure error suppression below-threshold.

The $p$-dependent finite-size offset $\delta(p)$ (Fig.~\ref{fig:deltap}), originates from the logarithmic and constant subleading corrections to the asymptotics of $\log(\mathbb{P}_{\mathrm{fail}})$ from the DPRE picture. Across the ordered regime, $\delta(p)$ remains positive, of order unity, and shows only a weak variation with $p$. The presence of a non-zero $\delta(p)$ underscores the energetic contributions from disorder beyond the simple linear scaling with $L$, and can be directly related to the critical amplitude of $\mathbb{P}_{\mathrm{fail}}$ at threshold, which is independent of $L$. Furthermore, the success of our error function ansatz for the finite-size scaling function (Eq.~\eqref{eq:FSS_general_erf_ansatz}), derived from a transformed Gaussian distribution of ground state energy and free energy costs, suggests that while the underlying energy landscape is complex and non-Gaussian in its full detail \cite{PhysRevB.75.174415}, the critical failure statistics are predominantly governed by fluctuations that are effectively Gaussian in nature after appropriate rescaling and shifting. This provides a powerful simplification for modeling near-threshold behaviour.

\subsection{Comparison to prior work}

Here we compare our results with existing literature. Dennis \emph{et al.} \cite{10.1063/1.1499754} established the mapping of surface codes to a 2D RBIM (and with measurement errors to a 3D random-plaquette gauge model) and identified the threshold as a phase transition. We instead use full post-selection to obtain a clean Ising model on the torus and derive closed-form expressions for $\mathbb{P}_{\mathrm{fail}}(p,L)$ for all $p,L$ in that setting, then transfer these insights to non-post-selected codes via an effective tension $\sigma_{\mathrm{eff}}(p)$. On the Nishimori line, Wang, Harrington and Preskill~\cite{WangChenyang2003Ctia} related the threshold to critical phenomena and provided a finite-size scaling analysis. The finite-size scaling ansatz in Ref.~\cite{WangChenyang2003Ctia} involved a Taylor expansion of the scaling function close to the point where the logical error probabilities cross. Our error function scaling ansatz captures the whole region where the logical error probability is changing rapidly with the physical error probability and was inspired by the distribution of domain wall free energy costs at threshold.

Watson and Barrett \cite{Watson_2014} identified a small-$p$ path-counting regime and a near-threshold scaling regime that inform overhead estimates. Our results recover these same limits quantitatively: (i) we reproduce the path-counting behaviour when $p\ll 1/L^{2}$; (ii) in the below-threshold regime our expression reduces asymptotically to their exponential-decay ansatz as it approaches the near-threshold regime, while giving a physical interpretation of the coefficients of exponential decay and (iii) we improve their scaling ansatz by deriving the scaling function from the full distribution of domain wall energy costs, which yields an explicit error function collapse with interpretable parameters. 

Beverland \emph{et al.} \cite{Beverland_2019} analyzed the tradeoff between square-lattice and rotated surface codes, emphasizing how combinatorial counting of low-weight logical errors and boundary condition effects govern performance at small $p$ while such boundary effects are not significant close to threshold. Our analysis extends this by deriving closed-form path-counting expressions for the surface code with a range of boundary conditions and identifying the precise validity window $p \ll 1/L^{2}$ for the path-counting approximation. 

Bravyi and Vargo \cite{PhysRevA.88.062308} used rare event Monte Carlo (splitting) methods to estimate $\mathbb{P}_{\mathrm{fail}}(p,L)$ for surface codes with holes up to $L\leq 20$, confirming $\mathbb{P}_{\mathrm{fail}}\sim e^{-\alpha(p)L}$, and providing a fit for the decay rate $\alpha(p)$. Our analysis replaces such empirical decay rate fits with formulas from the clean Ising mapping, identifying \(\alpha(p)=\beta\,\sigma_{\mathrm{eff}}(p)\) as a domain wall surface tension. Bravyi, Suchara, and Vargo~\cite{PhysRevA.90.032326} gave a polynomial-time MLD for independent bit/phase-flip noise via matchgate simulation (and an MPS-based approximation for depolarizing noise); our results are decoder-agnostic and provide parameter-interpretable fits for $\mathbb{P}_{\mathrm{fail}}(p,L)$ that can be applied to outputs of any MLD or MPD implementation.

Most recently, Xiao, Srivastava, and Granath \cite{Xiao2024exactresultsfinite} derived closed-form logical failure rates for the XY and XZZX surface codes under biased noise by choosing rotated boundaries that reduce a special disordered point to decoupled 1D Ising chains; our results are complementary, providing closed-form \(\mathbb{P}_{\mathrm{fail}}(p,L)\) for the unbiased, fully post-selected case and a physical identification \(\alpha(p)=\beta\,\sigma_{\mathrm{eff}}(p)\) that extends beyond a single bias-specific point.

\subsection{Realm of applicability and limitations}

Our four-regime framework applies broadly to topological stabilizer codes in two spatial dimensions, which have local checks and codimension-1 logical operators, under i.i.d. Pauli noise on the Nishimori line, where maximum-likelihood decoding maps to free energy differences in a classical spin model. Under these conditions, the decoding problem exhibits an order-disorder transition enabling the finite-size scaling we use. A more detailed explanation of the assumptions required for our framework to apply, along with further examples, and technical bounds on domain wall free energies are presented in Appendix~\ref{app:scope}.

While this work provides a comprehensive framework, certain limitations should be acknowledged. Our primary analysis has focused on $D=2$ surface codes under a bit-flip noise model, leveraging its direct mapping to Ising-type models. The extension of the exact analytical treatment of the fully post-selected case to more complex noise channels like depolarizing noise, which maps to the eight-vertex model, remains an open challenge, though our general four-regime structure is expected to persist. For the non-post-selected codes, our analysis relied on $T=0$ (MWPM) decoding and $T=T_{N}(p)$ optimal MLD only. The impact of finite-temperature decoding off the Nishimori line (e.g., as proposed recently in Ref.~\cite{wichette2025}, or other decoding schemes) should also be investigated.

Additionally, while our ans\"{a}tze demonstrate excellent agreement with numerical data, they are phenomenological in parts, and a more rigorous derivation from first principles of the random-bond Ising model for some of these forms (particularly the $p$-dependence of $\delta(p)$ or the precise form of the energy cost distribution leading to the error function scaling) would be desirable. Finally, the range of system sizes $L$ and the number of shots in our numerical simulations, while substantial, inherently limit the precision to which asymptotic scaling behaviours can be confirmed, especially at extremely low $p$.

\subsection{Outlook}

Future work should determine the finite-size scaling functions governing fault-tolerant logical operations such as lattice surgery~\cite{horsman2012surface}, building on recent statistical mechanical studies of lattice surgery-based logical state teleportation and its crossover scaling behaviour~\cite{ppng-vbqj}, and more broadly including the impact of boundary conditions~\cite{fowler2013accurate, bombin2023logical}, geometry~\cite{PhysRevA.76.012305, newman2020generating, bombin2023fault}, and topological features such as twist defects~\cite{bombin2010topological}. We expect that the underlying bulk scaling behaviour remains the same as in the defect-free case, with any deviations captured entirely by modified finite-size (or boundary) scaling functions. For instance, while the primary scaling variable $x=(p-p_{c})L^{1/\nu}$ would retain its form, the near-threshold failure probability $\mathbb{P}_{\mathrm{fail}}(p,L)$ would be described by a boundary-dependent scaling function. Thus, $\mathbb{P}_{\mathrm{fail}}(p\approx p_{c},L,B)\approx f_{B}((p-p_{c})L^{1/\nu})$, where $f_{B}(x)$ explicitly depends on the set of parameters $B$ characterizing the specific boundary conditions or geometry, distinguishing it from an idealized scaling function $f_{0}(x)$.

Beyond these scaling investigations, the exploration of new code constructions derived from statistical mechanical models exhibiting high-temperature transitions could promise new topological code constructions. Given the exact solvability of the eight-vertex model, future work could also explore extending the current statistical mechanics mapping to the surface code under depolarizing noise, which maps to the eight-vertex model in the fully post-selected limit. 

Furthermore, investigating the model's applicability to scenarios involving measurement errors, potentially through connections to random plaquette gauge models, or more complex circuit-level noise, also presents a direction for research. It would be of interest to settle the question of whether topological codes with measurement errors map to statistical mechanical models with first order or continuous phase transitions. This would require a careful consideration of the possibility that these transitions are showing \emph{pseudo-first-order} behaviour~\cite{Jin2012,Agrawal2025a}. In the case of a first order transition our finite-size scaling approach may not work at all system sizes, although in the examples we have studied in this work we were able to analyse the logical failure rate as a non-local order parameter through finite-size scaling.

While the preceding points outline directions for extending and refining the theoretical framework, the approach developed in this paper also holds immediate promise for analyzing specific, practical code families. For instance, an interesting application of the approach developed here is to families of bivariate bicycle codes~\cite{Bravyi2024} with fixed check structure \cite{chen2025anyontheorytopologicalfrustration,liang2025generalizedtoriccodestwisted}, as these codes satisfy the assumptions on which our analysis is founded. This could provide accurate and efficient estimates of optimal logical error rates in practical regimes for near-term quantum error correction with quantum low-density parity-check codes. Finally, for near-term quantum computing platforms that operate with $p>p_{c}$ yet remain in the near-threshold regime, our framework suggests a practical diagnostic: by fitting logical failure data to our finite-size scaling forms, one can extract a quantitative `distance-to-threshold' metric. This would enable experimental systems to benchmark their proximity to fault tolerance and track progress toward entering the below-threshold regime.

\begin{acknowledgements}
We thank Christopher Chubb, Yaodong Li and Nicholas O’Dea for helpful discussions. We also thank an anonymous referee for exceptionally thoughtful and detailed feedback, particularly with respect to the asymptotics of the distribution of domain wall energy costs in the random-bond Ising model. This work is supported by the Australian Research Council via the Centre of Excellence in Engineered Quantum Systems (EQUS) project number CE170100009, and by the ARO through the QCISS program W911NF-21-1-0007 and IARPA ELQ program W911NF-23-2-0223. 
DJW is supported by the Australian Research Council Discovery Early Career Research Award (DE220100625). 
\end{acknowledgements}

%

\appendix

\section{Free fermion sectors and spin partition functions on the torus}
\label{app:fermions-on-torus}

In this section, we outline how we evaluate the free fermion partition functions of the 2D Ising model on an $L\times L$ torus with twisted (periodic/antiperiodic) boundary conditions, and how these sectors are recombined to obtain the spin partition functions with $(pp)$, $(pa)$, $(ap)$, and $(aa)$ boundaries. Our presentation follows Refs.~\cite{Ming-ChyaWu_2002,Izmailian_2012}. The fermions acquire a mass away from criticality according to Eq.~\eqref{eq:mu}. For later use we record the lattice dispersion relation
\begin{equation}
    \omega_{\mu}(k)=\mathrm{arcsinh}\left(\sqrt{\sin^{2}(k) + 2\sinh^{2}(\mu)}\right).
\end{equation}
Labeling the fermion boundary conditions along the two cycles by $(\alpha,\gamma)\in\{0,\tfrac{1}{2}\}^{2}$, where $0$ denotes periodic and $\tfrac{1}{2}$ antiperiodic, the partition function in sector $(\alpha,\gamma)$ is
\begin{equation}
    Z_{\alpha,\gamma}(\mu)=\prod_{n=0}^{L-1}2\left| \sinh\left[L\omega_{\mu}\left(\frac{\pi(n+\alpha)}{L}\right) + i\pi\gamma\right]\right|.
\end{equation}
The spin partition functions for the four spin boundary conditions on the torus are linear combinations of these fermionic sectors. Writing $p$ for periodic and $a$ for antiperiodic spin boundary conditions, we have
\begin{align}
    Z_{pp}&=\frac{(2\sinh(2J\beta))^{\frac{L^{2}}{2}}}{2}\biggl( Z_{1/2,1/2}(\mu) + Z_{0,1/2}(\mu)\nonumber\\
    &+ Z_{1/2,0}(\mu) - \mathrm{sgn}\left(\frac{T-T_{c}}{T_{c}}\right)Z_{0,0}(\mu) \biggl),\nonumber\\
    Z_{pa}&=\frac{(2\sinh(2J\beta))^{\frac{L^{2}}{2}}}{2}\biggl( Z_{1/2,1/2}(\mu) + Z_{0,1/2}(\mu)\nonumber\\
    &- Z_{1/2,0}(\mu) + \mathrm{sgn}\left(\frac{T-T_{c}}{T_{c}}\right)Z_{0,0}(\mu)\biggl),\nonumber\\
    Z_{ap}&=\frac{(2\sinh(2J\beta))^{\frac{L^{2}}{2}}}{2}\biggl( Z_{1/2,1/2}(\mu) - Z_{0,1/2}(\mu)\nonumber\\
    &+ Z_{1/2,0}(\mu) + \mathrm{sgn}\left(\frac{T-T_{c}}{T_{c}}\right)Z_{0,0}(\mu)\biggl),\nonumber\\
    Z_{aa}&=\frac{(2\sinh(2J\beta))^{\frac{L^{2}}{2}}}{2}\biggl( -Z_{1/2,1/2}(\mu) + Z_{0,1/2}(\mu)\nonumber\\
    &+ Z_{1/2,0}(\mu) + \mathrm{sgn}\left(\frac{T-T_{c}}{T_{c}}\right)Z_{0,0}(\mu)\biggl).
\end{align}

\section{Surface code failure rates in path-counting regime}\label{app:path_count}

In this section, we derive the leading-order expressions for the logical failure rate of various surface codes in the path-counting regime by counting the number of most likely error paths. Our analysis follows that of Beverland \emph{et al.}~\cite{Beverland_2019}, who demonstrated a significant difference in this path-counting regime between the rotated and unrotated surface codes due to a larger number of low-weight error chains in the rotated version. This highlights the substantial effect of boundary conditions on logical error rates when the system is far from the thermodynamic limit. This behaviour contrasts with the critical regime near-threshold, where they showed that both code geometries exhibit nearly identical behaviour, reflecting the universality of bulk properties in topological codes where boundary effects are diminished.

\subsection{Counting errors}

Consider path counting following Dennis \emph{et al.}~\cite{10.1063/1.1499754} and Beverland \emph{et al.}~\cite{Beverland_2019}. Specifically, we can take the upper bound from Eq.~(22) of Beverland \emph{et al.}~\cite{Beverland_2019}
\begin{align}
\mathbb{P}_{\mathrm{fail}}&\leq (1 - p)^n \sum_{l = L}^{n} N_{\text{con}}(l)\times\nonumber\\
&\sum_{u = \lceil l/2 \rceil}^{l} \sum_{v = 0}^{n - l} 
\binom{l}{u} \binom{n - l}{v} \left[ \frac{p}{1 - p} \right]^{u + v},
\end{align}
where $N_{\mathrm{con}}(l)$ is the number of non-contractible closed paths on the lattice. The sum over $u$ reflects the number of ways of removing $l-u$ errors from the closed path to leave an uncorrectable error. The sum over $v$ reflects additional errors separate from the error chain that are assumed not to affect performance of the decoder. This is an upper bound because the sum over $v$ will include a very low probability contribution of additional non-contractible errors and thus there is some double counting of error configurations in this sum. However for low enough $p$ and odd $L$ this double counting should be negligible and the upper bound should be a good estimate of the logical error probability. The expressions for even $L$ will typically result in an extra factor of a half since there are error configurations with weight $L/2$ where the decoder fails only half the time. Specifically if $p \sim 1/L$ and $n = L^{2}$, then there are on average $L$ errors in the code. There are ${L^{2} \choose L}\sim (L-1)^{L}e^{L}$ such error configurations of which only $2L$ are non-contractible closed paths on the torus. Thus we expect this formula to be a good estimate of the logical error probability at least when $p \lesssim 1/L$. The estimate of the binomial coefficient here is Stirling’s approximation for large $L$. Unless $p \ll 1/n$ the prefactor $(1-p)^{n} \sim e^{-np}$ is not close to one. But this formula can be further analysed for moderately low $p \lesssim 1/L$ by performing the sum over errors $v$ that are separate from the nontrivial error chain. Using Eq.~(23) of Beverland \emph{et al.}~\cite{Beverland_2019} we find the estimate
\begin{equation}
\mathbb{P}_{\mathrm{fail}} \simeq \sum_{l = L}^{n} N_{\text{con}}(l)(1 - p)^l \sum_{u = \lceil l/2 \rceil}^{l} 
\binom{n}{u} \left[ \frac{p}{1 - p} \right]^u.
\end{equation}
Below, we analyse this formula to find the leading order logical error probability as follows:
\begin{equation}
\mathbb{P}_{\mathrm{fail}}\simeq N_{\text{min}} p^{\lceil L/2 \rceil} + N_1 p^{\lceil L/2 \rceil + 1} + \dots
\end{equation}
This formula requires counting the minimum weight errors that cause decoder failure and is expected to be good so long as \( p \ll N_{\text{min}}/N_1 \). These numbers $N_{\mathrm{min}},N_{1}$ depend strongly on the boundary conditions.

\subsection{Torus, odd distance}

The smallest possible logical operator $l=d$ corresponds to straight lines across the code, of which there are $N_{\mathrm{con}}(L) = 2L$. The factor of 2 corresponds to horizontal and vertical lines. These straight lines lead to an error of weight $\lceil L/2\rceil$ that
is incorrectly decoded. These uncorrectable errors correspond to removing $\lfloor L/2
\rfloor$ errors from such a lowest weight logical operator. Thus
\begin{equation}
N_{\text{min}} = 2L (1 - p)^{\lfloor L/2 \rfloor} \binom{L}{\lfloor L/2 \rfloor}.
\end{equation}
This gives us the lowest order path counting estimate for the logical error probability. 

The second shortest logical operators are lines across the code that have $l = L + 2$ and have 2 kinks added to one of the $2L$ straight lines. There are $L(L-1)/2$ possible locations for the kinks, and 2 possible orientations for the kink, so that $N_{\mathrm{con}}(L + 2) = (2L)L(L-1) = 2(L^{3}-L^{2})$. This curve leads to a decoder error for a Pauli of weight $\lceil L/2\rceil + 1$ once we have removed $\lfloor L/2\rfloor + 1$ errors from the logical operator. We can also remove $\lfloor L/2\rfloor-1$ errors from the straight line to find another error configuration with this weight that results in decoder failure and so
\begin{align}
N_1 = {} & 2(L^3 - L^2)(1 - p)^{\lfloor L/2 \rfloor + 1} \binom{L + 2}{\lfloor L/2 \rfloor + 1} \nonumber\\
& + 2L (1 - p)^{\lfloor L/2 \rfloor - 1} \binom{L}{\lfloor L/2 \rfloor - 1}.
\end{align}
Comparing these two in the case of large d we find that we need
\[
p \ll \frac{1}{4L^2},
\]
this tells us the range of validity of the path counting regime, with higher order terms starting to dominate at large physical error probabilities. Given this range of $p$ we have $(1-p)^{\lfloor L/2\rfloor} \simeq 1$ so we can neglect these factors in this regime. However note that it is important that we do not have a factor $(1-p)^{n}$ here since that is not necessarily close to one in this regime.

\subsection{Torus, even distance}\label{subsec:toruseven}

For even distance we again care about $l = L$ strings. Remove $L/2$ errors results in an error chain of weight $L/2$ for which the decoder will fail $1/2$ the time. So for even distance we find
\begin{equation}
N_{\text{min}} = L (1 - p)^{\lfloor L/2 \rfloor} \binom{L}{\lfloor L/2 \rfloor}.
\end{equation}

It is straightforward to calculate $N_1$ for this case also and determine that the regime of validity for this logical error probability estimate is $p\ll 1/4L^2$.

\subsection{Planar, even distance}

With the planar code patch, there are now half as many length $d$ error chains since only horizontal chains exist. However logical operators do not need to satisfy periodic boundary conditions, so there now exist lines with length $L+1$ that have a single kink and result in a logical operation. Since $L$ is even we can remove $L/2 + 1$ errors from the line to leave behind a configuration of $L/2$ errors. We will assume that this is a correctable chain. In that case for the even planar code the minimum weight errors result from straight lines of length $L$ and we have
\begin{equation}
N_{\text{min}} = \frac{1}{2} L (1 - p)^{\lfloor L/2 \rfloor} \binom{L}{\lfloor L/2 \rfloor}.
\end{equation}
For the longer error chain we need to consider the contribution of both $l = L + 1$ chains with $L/2$ errors removed, and $l = L + 2$ chains with $L/2 + 1$ errors removed, which will lead to decoder failure half the time. The latter case will be the dominant contribution due to the larger number of these error configurations. The $l = L+1$ chains have a single kink, this kink can go in one of two orientations, so there are $L(2L) = 2L^{2}$ of these. The $l = L + 2$ chains have two kinks, but unlike the periodic case there are 4 possible orientations of the pair of kinks, so there are $2L(L-1)$ kink configurations for each of the $L$ $l = L$ chains. Thus there are $2(L^{3}-L^{2})$ of these error configurations. Finally as before we can remove $L/2-1$ errors from the $l = L$ chains. So we have
\begin{align}
N_1 = {} & 2(L^3 - L^2)(1 - p)^{\lfloor L/2 \rfloor + 1} \binom{L + 2}{\lfloor L/2 \rfloor + 1} \nonumber\\
& + 2L^2 (1 - p)^{\lfloor L/2 \rfloor} \binom{L + 1}{\lfloor L/2 \rfloor} \nonumber\\
& + L (1 - p)^{\lfloor L/2 \rfloor - 1} \binom{L}{\lfloor L/2 \rfloor - 1}.
\end{align}
The first of these terms is dominant at large $L$ and again tells us that we need $p \ll 1/4L^{2}$.

\subsection{Planar, odd distance}

For odd distance we can remove $\lceil L/2\rceil$ errors from the $l = L + 1$ chains to leave $\lceil L/2\rceil$ errors, and the decoder will fail half the time for these. This will mean that the logical failure rate is higher for the odd size surface code by a
factor of $L$. This gives us
\begin{align}
N_{\text{min}} = {} & L^2 (1 - p)^{\lfloor L/2 \rfloor} \binom{L + 1}{\lceil L/2 \rceil} \nonumber\\
& + \frac{1}{2} L (1 - p)^{\lfloor L/2 \rfloor} \binom{L}{\lfloor L/2 \rfloor}.
\end{align}
Again we can drop the factors $(1-p)$ in the relevant regime.

This analysis suggests that the different combinatorial factors for the number of low-weight error chains, particularly the emergence of $l=L+1$ error chains in the planar code with open boundaries, could be the origin of the pronounced even-odd distance effect observed in surface codes. This effect, where codes with an odd distance $L$ exhibit a higher logical error rate than those with an even distance, has been noted in numerical simulations starting with the work of Wang, Harrington, and Preskill \cite{WangChenyang2003Ctia} and persists even for relatively large code sizes close to the threshold.

\section{Asymptotics of post-selected surface tension and non-post-selected effective tension}\label{app:asymptotics}

In this section, we describe the asymptotic forms for the tensions of both fully post-selected and non-post-selected surface codes. We start from the decay rate definition
\begin{equation}
\alpha(p)=\beta\sigma(p)\;=\;-\lim_{L\to\infty}\frac{1}{L}\log \mathbb{P}_{\mathrm{fail}}(p,L),
\label{eq:F-alpha-def}
\end{equation}
and use the Nishimori conditions
\( \beta J=\frac{1}{2}\log\frac{1-p}{p}\) 
throughout.

\subsection{Fully post-selected surface codes}

For the clean 2D Ising model ($T<T_c$) the domain wall free energy per unit length splits into an energetic piece $2J$ from broken bonds and an entropic reduction that vanishes as $\beta\to\infty$:
\begin{equation}
  \sigma(\beta) = \overbrace{2J}^\text{energetic} - \overbrace{\frac{1}{\beta}\log\coth(\beta J)}^{\text{entropic}}
\end{equation}
The second term should not be confused with the finite-size capillary-wave correction $-\frac{1}{2}\log L$ that arises from the roughening of the interface (see, e.g., Ref.~\cite{PhysRevE.90.012128}). Along the Nishimori line, $\coth(\beta J)=\frac{1}{1-2p}$, so we can write
\begin{align}
  \alpha(p)=\beta\sigma(p) = \overbrace{\log\left(\frac{1-p}{p}\right)}^\text{energetic} + \overbrace{\log(1-2p)}^\text{entropic}.\label{eq:path_count_energetic_entropic}
\end{align}
The path-counting approximation
$\mathbb{P}_{\mathrm{fail}}\approx 2L\!\left(\tfrac{p}{1-p}\right)^{\!L}$
is valid only for $p\ll 1/L$ (i.e., $pL\to 0$ as $L\to\infty$; see Sec.~\ref{subsec:path_counting_postselect}),
whereas the decay rate
\[
\alpha(p)=-\lim_{L\to\infty}\frac{1}{L}\log \mathbb{P}_{\mathrm{fail}}(p,L)
\]
holds for $p$ fixed as $L\to\infty$. These limits do not commute, so the path-counting
form should not be used to determine $\alpha(p)$. Indeed, if one enforces a fixed
$pL=\mathcal{O}(1)$ by taking $p=c/L$, then
\[
-\frac{1}{L}\log\Big[2L\left(\frac{p}{1-p}\right)^{L}\Big]
= \log L - \log c + \mathcal{O}(1),
\]
which diverges as $\log L$. Instead, using the clean domain wall free energy (Eq.~\eqref{eq:path_count_energetic_entropic}) one has $\alpha(p)=\log\!\frac{(1-p)(1-2p)}{p}=\log\frac{1}{p}+\mathcal{O}(p)$, so along the Nishimori line
\[
\sigma(p)=\frac{\alpha(p)}{\beta}
=2J\left[1+\mathcal{O}\left(\frac{1}{\log(1/p)}\right)\right],
\]
i.e., $\sigma(p)\to 2J$ with $1/\log(1/p)$ corrections as $p\to 0$.

Near the clean critical point \(p_c=\frac{1}{2+\sqrt{2}}\), a first-order expansion of Eq.~\eqref{eq:path_count_energetic_entropic} gives
\begin{align}
\alpha(p)&=-4(1+\sqrt{2})(p-p_{c})+\mathcal{O}\left((p-p_{c})^2\right),
\label{eq:F-alpha-linear}
\end{align}
so \( \beta\sigma\) vanishes linearly as \(p\to p_c\), matching the finite-size scaling analysis with $\nu=1$. Dividing by the thermodynamic $\beta$ according to the Nishimori conditions then yields an explicit leading form for the surface tension,
\begin{equation}
\sigma(p)=\frac{-8(1+\sqrt{2})\,(p-p_{c})}{\log\frac{1-p}{p}},
\qquad p\to p_c.
\label{eq:F-sigma-clean-nearpc}
\end{equation}

\subsection{Non-post-selected surface codes}

The path-counting approximation for the non-post-selected case (see Appendix~\ref{app:path_count}) is valid only for $p\ll 1/L^{2}$, i.e., $pL^{2}\to 0$ as $L\to\infty$, whereas the decay rate holds $p$ fixed. These limits do not commute. Enforcing a fixed $pL^{2}=\mathcal{O}(1)$ by taking $p=c/L^{2}$ in the path-counting form
\[
\mathbb{P}_{\mathrm{fail}}\approx L\binom{L}{L/2}p^{L/2}
\]
and applying Stirling's approximation for $\log\binom{L}{L/2}=L\log 2-\tfrac12\log(\pi L/2)+\mathcal{O}(1)$ yields
\begin{align*}
-\frac{1}{L}\log \mathbb{P}_{\mathrm{fail}}(L,p=c/L^{2})
&= \log L - \log 2 - \tfrac12\log c + \mathcal{O}(1),
\end{align*}
which similarly diverges as $\log L$. Thus the path-counting expression is incompatible with the decay-rate limit and cannot be used to determine $\alpha(p)$. With the Nishimori conditions satisfied, taking $p\to 0$ sends $\beta J=\frac{1}{2}\log\frac{1-p}{p}\to\infty$ while the bond randomness vanishes, so the RBIM reduces to the clean $T=0$ Ising model; consequently,
\begin{equation}
    \lim_{p\to 0}\sigma(p)=2J.
\end{equation}
The limit above refers to the microscopic clean Ising surface tension. In the non-post-selected decoding problem, the path-counting approximation effectively enforces domain walls of length $L/2$ (assuming even $L$; see Sec.~\ref{subsec:toruseven}). Modeling the failure probability as $\mathbb{P}_{\mathrm{fail}}\sim \exp[-\beta\sigma_{\mathrm{eff}}(p)L/2]=\exp[-\alpha(p)L]$ and comparing with the decay-rate definition (Eq.~\eqref{eq:F-alpha-def}) identifies an effective tension $\sigma_{\mathrm{eff}}(p)=\sigma(p)/2$. Hence $\sigma_{\mathrm{eff}}\to J$ as $p\to 0$ (equal to $1$ when $J=1$), consistent with the small-$p$ limits reported by Beverland \emph{et al.}~\cite{Beverland_2019}.

From the RBIM mapping, the near-threshold finite-size scaling implies \(\mathbb{P}_{\mathrm{fail}}(p,L)=f\left( (p-p_c)L^{1/\nu}\right)\) with \(\nu\neq 1\) (empirically \(\nu\approx 1.5\) for bit-flip noise). Consistency with this scaling and the decay-rate definition implies that, as \(p \to p_c\) from below,
\begin{align}
\alpha(p)&=\beta\sigma_{\mathrm{eff}}(p)\propto |p-p_{c}|^{\nu},
\\
\Rightarrow\quad
\sigma_{\mathrm{eff}}(p) &= \frac{C\,|p-p_{c}|^{\nu}}{\frac{1}{2J}\log\frac{1-p}{p}},
\label{eq:F-sigma-nps-nearpc}
\end{align}
for some nonuniversal amplitude \(C\). This viewpoint was introduced by Watson and Barrett \cite{Watson_2014}, who proposed the finite-size scaling ansatz \( \mathbb{P}_{\mathrm{fail}}(p,L)= Ae^{-a|p-p_{c}|^{\nu}L} \), for $p<p_{c}$ which both guarantees exponential decay of logical failure rates below-threshold and yields data collapse under the rescaling by $x=(p-p_c)L^{1/\nu}$.

\section{Derivation of capillary wave corrections to surface tension}\label{app:capillary}

In this section, we sketch a derivation of the capillary wave corrections to the free energy cost of a domain wall in the 2D Ising model at finite thermodynamic beta ${\beta=\frac{1}{k_{B}T}}$ in the ferromagnetic (ordered) phase. In the thermodynamic limit the free energy cost is given by Onsager’s result, $\Delta F_{\mathrm{bulk}}(\beta)=\sigma(\beta)L$, with $\sigma(\beta)=\big(2J-\frac{1}{\beta}\log\coth(J\beta)\big)$ \cite{PhysRev.65.117}. For finite systems the interface fluctuates, and these capillary waves modify the free energy by an additive logarithmic term \cite{doi:10.1142/S0129183192000531}.

In the capillary wave model the fluctuating interface is described by a height function $h(x)$ along the coordinate $x$ (taken along the mean direction of the interface). In this case, the boundary tension becomes a function of the slope of the interface: $\sigma(\theta,\beta)$. The local slope of the interface is given by
\begin{equation}
    \theta(x)=\arctan\!\Bigl(\frac{dh}{dx}\Bigr),
\end{equation}
so that the differential arc length is
\begin{equation}
    dl=\sqrt{1+\Bigl(\frac{dh}{dx}\Bigr)^2}\,dx.
\end{equation}
The macroscopic energy of the interface is then obtained by integrating the angle-dependent surface tension over this arc length
\begin{equation}
    E = \frac{1}{\beta}\int_{0}^{L} dx\, \sigma\Bigl[\arctan\Bigl(\frac{dh}{dx}, \beta\Bigr)\Bigr]\, \sqrt{1+\Bigl(\frac{dh}{dx}\Bigr)^2}\,.
\end{equation}
For nearly flat interfaces (where $\frac{dh}{dx}$ is small), one can expand both 
$\sigma\Bigl[\arctan\!\bigl(\frac{dh}{dx}\bigr), \beta\Bigr]$ and the square-root factor in powers of $\frac{dh}{dx}$. Keeping terms up to second order yields
\begin{align}
    &\sigma\Bigl[\arctan\!\Bigl(\frac{dh}{dx}, \beta\Bigr)\Bigr]\sqrt{1+\Bigl(\frac{dh}{dx}\Bigr)^2}\nonumber\\
    &\approx \sigma(0, \beta) + \sigma'(0, \beta)\frac{dh}{dx} + \frac{1}{2}\kappa(\beta) \Bigl(\frac{dh}{dx}\Bigr)^2\,,
\end{align}
where the surface stiffness coefficient is $$\kappa(\beta)\equiv\sigma(0, \beta)+\sigma''(0, \beta).$$ 
The first term, $\sigma(0)$ contributes a constant reference energy proportional to the total interface length, while the linear term in $\frac{dh}{dx}$ contributes only at the boundaries (which we may neglect under periodic boundary conditions). Thus, the fluctuating part of the interface energy is given by
\begin{equation}
    E_{cw}\beta = \frac{1}{2}\kappa(\beta)\int_{0}^{L} dx\, \Bigl(\frac{dh}{dx}\Bigr)^{2}.
\end{equation}
This is the standard capillary wave Hamiltonian in 2D \cite{doi:10.1142/S0129183192000531,PhysRevE.90.012128}. To evaluate the partition function, we decompose the height fluctuations into Fourier modes
\begin{equation}
    h(x)=\sum_{q} \tilde{h}_q\,e^{iqx}\,,
\end{equation}
with wave vectors $q=2\pi n/L$ for a system of length $L$. We therefore have
\begin{align}
    \frac{dh}{dx}&= \sum_{q}iq\tilde{h}_{q}e^{iqx},\nonumber\\
    \left(\frac{dh}{dx}\right)^{2}&=\sum_{q}q^{2}|\tilde{h}_{q}|^{2}.
\end{align}
Substituting this expansion into the energy yields
\begin{align}
    E_{cw}\beta &= \frac{\kappa(\beta)}{2}\int_{0}^{L} dx \sum_{q} q^{2}|\tilde{h}_{q}|^{2}\nonumber\\
    &= \frac{\kappa(\beta)L}{2}\sum_{q} q^{2}|\tilde{h}_{q}|^{2}.
\end{align}
Because the Hamiltonian is quadratic, the partition function factors into a product over independent Gaussian integrals. The Gaussian integral below is taken only over the nonzero Fourier modes of the interface height field. The $q=0$ mode corresponds to the collective translation of the interface and is restored separately as a factor of $L$ for a free interface on the torus. We evaluate these integrals as
\begin{align}
    Z_{cw}&=\prod_{q\neq0}\int d\tilde{h}_q\, \exp\!\left[-\frac{\kappa(\beta)L}{2}\,q^2\,|\tilde{h}_q|^2\right]\nonumber\\
    &= \prod_{q\neq 0}\sqrt{\frac{\pi}{\frac{\kappa(\beta) L}{2}q^{2}}}\nonumber\\
    &= \prod_{n\neq 0}\frac{\sqrt{L}}{\sqrt{2\pi\kappa} n}, 
\end{align}
where we have evaluated the Gausssian integrals. This product diverges, so we apply zeta-function regularization to obtain a physically sensible result. First we split this product into the positive and negative modes separately, which we can rewrite as
\begin{equation}
    Z_{cw} = \prod_{n=1}^\infty \frac{\sqrt{L}}{\sqrt{2\pi\kappa}\, n} \times \prod_{n=1}^\infty \frac{\sqrt{L}}{\sqrt{2\pi\kappa}\, (-n)}\,.
\end{equation}
Next we can express the total partition function for the capillary wave fluctuations as
\begin{equation}
    Z_{cw} = \prod_{n=1}^\infty \left(\frac{\sqrt{L}}{\sqrt{2\pi\kappa}\, n}\right)^2 \times \prod_{n=1}^\infty (-1)\, ,
\end{equation}
where the infinite product of $-1$ over all positive $n$ leads to an undetermined overall phase. However, we know that the partition function must be positive, so we take the absolute value of this equation, obtaining
\begin{equation}
    |Z_{cw}|= \prod_{n=1}^\infty \left(\frac{\sqrt{L}}{\sqrt{2\pi\kappa}\, n}\right)^2.
\end{equation}
We can rewrite this expression as
\begin{equation}
    |Z_{cw}|=\prod_{n=1}^\infty \left(\frac{\sqrt{L}}{\sqrt{2\pi\kappa}\, n}\right)^2 = \left(\frac{L}{2\pi\kappa}\right)^{\sum_{n=1}^\infty 1}\,\prod_{n=1}^\infty \frac{1}{n^2}\,.
\end{equation}
Now we apply zeta function regularization, first noting $\sum_{n=1}^\infty 1 = \zeta(0) = -\frac{1}{2}$. Next we have $\prod_{n=1}^\infty n = e^{-\zeta'(0)} = \sqrt{2\pi}\,$,
such that $\prod_{n=1}^\infty n^2 = 2\pi\,$. Using the regularized values, we finally obtain
\begin{equation}
    |Z_{cw}|=\left(\frac{L}{2\pi\kappa}\right)^{-\frac{1}{2}}\frac{1}{2\pi}=\frac{1}{\sqrt{L}}\sqrt{\frac{\kappa}{2\pi}}
\end{equation}
On an $L\times L$ torus the domain wall can be translated to any of the $L$ transverse positions with the same energy. Summing over these degenerate placements multiplies the partition function by $L$:
\begin{equation}
    Z^{\mathrm{free}}_{cw} = L|Z_{cw}| 
= \sqrt{L}\sqrt{\frac{\kappa}{2\pi}}.
\end{equation}
Finally the free energy difference (remembering the energies are normalized in units of $\frac{1}{\beta}$ as above) is given by
\begin{equation}
    \Delta F^{\mathrm{free}}_{\mathrm{cw}} = -\frac{1}{2}\log\left(L\right) - \log\left(\sqrt{\frac{\kappa}{2\pi}} \right).
\end{equation}
In our convention the bare free energy cost of a rigid interface is given by $\Delta F_{\mathrm{bulk}}(\beta)=\sigma(\beta)L$. Adding the capillary-wave contribution for a free interface gives
\begin{equation}\label{eq:capillary_wave_full}
    \Delta F(L,\beta)=\sigma(\beta)L-\frac{1}{2}\log(L)-\log\left(\sqrt{\frac{\kappa}{2\pi}}\right).
\end{equation}
In practice, we absorb the logarithmic term in the stiffness coefficient, $\log\left(\sqrt{\frac{\kappa}{2\pi}}\right)$ into an overall constant. That is, we write $\Delta F(L,\beta)=\sigma(\beta)L-\frac{1}{2}\log(L) + \mathrm{const}$, and then neglect the constant term when comparing to numerical results. This approximation is justified both by the fact that the variation in $\kappa(\beta)$ is small over the temperature range of interest and by the excellent numerical agreement observed when using the truncated expression. Thus, we arrive at the approximate expression for the free energy cost of a domain wall:
\begin{equation}
    \Delta F(L,\beta)=(2J-\frac{1}{\beta}\log\coth(J\beta))L - \frac{1}{2}\log(L).
\end{equation}

\section{Near-threshold expansion of fully post-selected toric code and the critical slope of the scaling function}\label{app:near_threshold_expansion}

We follow Ref.~\cite{Ivashkevich_2002} to derive the near-threshold expansion of the logical failure probability \( \mathbb{P}_{\mathrm{fail}} \) of a fully post-selected toric code on an \(L\times L\) torus to first order in the mass parameter $\mu$ and express the coefficients in terms of the Dedekind eta and Jacobi theta functions.

With \(Z_{\alpha,\beta}\) the fermionic partition functions at twist \((\alpha,\beta)\in\{0,\tfrac12\}^2\), one finds
\begin{align}
\mathbb{P}_{\mathrm{fail}}
=&\frac{1}{2}+
\nonumber 
\\ 
&\frac{\mathrm{sgn}\,Z_{0,0}(\mu)}{Z_{1/2,1/2}(\mu)+Z_{0,1/2}(\mu)+Z_{1/2,0}(\mu)+\mathrm{sgn}\,Z_{0,0}(\mu)},
\label{Pfail_appendix}
\end{align}
where $\mathrm{sgn}=\mathrm{sgn}(T-T_{c})$. Near the critical point \(\mu=0\), it can be shown that
\begin{align}
Z_{\alpha,\beta}(\mu)&=Z_{\alpha,\beta}(0)+\frac{\mu^2}{2}\,Z^{\prime\prime}_{\alpha,\beta}(0)+\cdots, \quad (\alpha,\beta)\neq(0,0), \\
Z_{0,0}(\mu)&=\lvert \mu\rvert\,Z^\prime_{0,0}(0)+\frac{\lvert \mu\rvert^3}{6}\,Z^{\prime\prime\prime}_{0,0}(0)+\cdots.
\end{align}
Defining \(S_0:=Z_{1/2,1/2}(0)+Z_{0,1/2}(0)+Z_{1/2,0}(0)\), a standard linearisation of Eq.~\eqref{Pfail_appendix} gives
\begin{equation}\label{Pfail_appendix_mu}
\mathbb{P}_{\mathrm{fail}}
=\frac{1}{2}+\frac{Z^\prime_{0,0}(0)}{S_0}\,\mu+\mathcal{O}(\mu^2),
\end{equation}
where we absorbed \(\mathrm{sgn}\,Z_{0,0}\) into the signed \(\mu\) so that the first correction is odd in \(\mu\). Following Eqs.~(16) and (17) of Ref.~\cite{Ivashkevich_2002}, evaluated at aspect ratio $\rho=1$ and equal lattice dimensions $L=M$ we have
\begin{align}
\log Z_{\alpha,\beta}(0)
&= \frac{2\gamma}{\pi}L^2 + \log\left|\frac{\theta_{\alpha,\beta}(i)}{\eta(i)}\right| + \mathcal{O}\left(\frac{1}{L^2}\right),
\nonumber \\
& \hspace{3.85cm} (\alpha,\beta)\neq(0,0),\\
\log Z^\prime_{0,0}(0)
&= \frac{2\gamma}{\pi}L^2 + \tfrac12\log(8L^2)+ 2\log|\eta(i)| + \mathcal{O}\left(\frac{1}{L^2}\right).
\end{align}
Exponentiating and taking the ratio, the common exponential factor cancels, giving
\begin{align}
\frac{Z^\prime_{0,0}(0)}{S_0}
=& \sqrt{8}\,L\,
\frac{|\eta(i)|^{3}}{\,|\theta_{1/2,1/2}(i)|+|\theta_{0,1/2}(i)|+|\theta_{1/2,0}(i)|}
\nonumber \\
&+\mathcal{O}\left(\frac{1}{L}\right).
\end{align}
Substituting into Eq.~\ref{Pfail_appendix_mu} yields
\begin{align}\label{eq:Pfail_muL}
\mathbb{P}_{\mathrm{fail}}
\simeq & \frac12
+\frac{\sqrt{8}\mu L\;|\eta(i)|^{3}}{\,|\theta_{1/2,1/2}(i)|+|\theta_{0,1/2}(i)|+|\theta_{1/2,0}(i)|}
\nonumber \\
&
+\mathcal{O}(\mu^2)+\mathcal{O}\left(\frac{1}{L}\right),
\end{align}
Using the scaling identification $\mu L=-(2+\sqrt{2})x$ (see Eq.~\eqref{eq:mu_x_relation}), Eq.~\eqref{eq:Pfail_muL} shows
\begin{align}
&\left.\frac{d\,\mathbb{P}_{\mathrm{fail}}}{dx}\right|_{x=0}\nonumber\\
&=-(2+\sqrt{2})\sqrt{8}\frac{|\eta(i)|^{3}}{|\theta_{1/2,1/2}(i)|+|\theta_{0,1/2}(i)|+|\theta_{1/2,0}(i)|}.
\end{align}
The proportionality constant is fully determined by the special function values at $\tau=i$, which we numerically evaluate and plot in Fig.~\ref{fig:data_collapse}. The magnitude of the critical slope is
\begin{align}
\left|f'(0)\right|&=(2+\sqrt{2})\sqrt{8}
\frac{|\eta(i)|^{3}}{|\theta_{1/2,1/2}(i)|+|\theta_{0,1/2}(i)|+|\theta_{1/2,0}(i)|}\nonumber\\
&\approx 1.503,
\end{align}
where the overall sign depends on the convention chosen for the signed mass parameter $\mu$.

\section{Supplementary analysis of near-threshold scaling of non-post-selected surface codes}\label{app:supplementary_scaling}

This section provides additional technical details and supporting results for Sec.~\ref{subsec:nearthresh_nopostselect} of the main text. We present the full derivation of the scaling form for the domain wall energy distribution, numerical evidence for convergence of the rescaled distributions to a universal envelope, asymptotic constraints on the scaling function, and a quantitative comparison of alternative finite-size scaling ans\"{a}tze. These results provide justification for the modeling choices and parameter estimates reported there.

\subsection{Scaling form for the domain wall energy distribution}\label{app:scaling_form}

In Sec.~\ref{subsec:nearthresh_nopostselect} we proposed that the distribution of domain wall ground state energy costs at $T=0$ obeys a finite-size scaling form near the threshold. Here we present the derivation of that form in full, starting from the assumption that the logical failure probability $\mathbb{P}_{\mathrm{fail}}$ depends only on the scaling variable $x=(p-p_{c})L^{1/\nu}$. We therefore propose that the distribution takes the form \cite{PhysRevB.38.386,DAHuse_1987,DSFisher_1987}
\begin{equation}
    \mathbb{P}(\Delta E; p,L) = \frac{1}{\Delta E_0(L)} G\left( \frac{\Delta E}{\Delta E_0(L)}, (p - p_c)L^{\frac{1}{\nu}} \right),
\end{equation}
\begin{equation}
    \Delta E_0(L) \sim L^{\theta} \mathcal{F}((p - p_c)L^{1/\nu}),
\end{equation}
with $G$ the scaling function, $\Delta E_0(L)$ representing the characteristic energy scale and $\mathcal{F}$ capturing subleading corrections away from ${p=p_{c}}$. We identify $\Delta E_{0}(L)$ with the mean of the distribution, $\Delta E_{0}(L)\coloneq \langle \Delta E(p,L)\rangle$. At the critical point ($p=p_{c}$), the scaling variable $x=0$, so $\mathcal{F}$ is a constant, and the characteristic energy scales as $\Delta E_{0}(L)\sim L^{\theta}$. Changing variables via $u=\Delta E/L^{\theta}$, the failure probability may be rewritten as
\begin{align}
    \mathbb{P}_{\mathrm{fail}} &= \int_{-\infty}^{0} d\Delta E \, \frac{1}{L^\theta} G\left( \frac{\Delta E}{L^\theta}, (p - p_c)L^{\frac{1}{\nu}} \right)\nonumber\\
    &= \int_{-\infty}^{0} du \, G\left( u, (p - p_c)L^{\frac{1}{\nu}} \right).
\end{align}

Since $\mathbb{P}_{\mathrm{fail}}$ is known to be a universal function of ${x=(p-p_{c})L^{1/\nu}}$, it follows that the scaling function $G(u,x)$ must depend on $p$ and $L$ only through the scaling variable $x$. In particular, at criticality $(p=p_{c})$, the domain wall energy distribution assumes the form
\begin{equation}\label{eq:spacing}
    \mathbb{P}(\Delta E; p_{c}, L) = \frac{1}{L^{\theta}} G\left( \frac{\Delta E}{L^{\theta}}, 0 \right).
\end{equation}

A key subtlety arises from the discrete nature of $\Delta E$ due to the bimodal coupling distribution, as noted in Ref.~\cite{PhysRevLett.91.087201}. For any finite $L$, the allowed values of $\Delta E$ are quantized, and hence the PMF $\mathbb{P}(\Delta E; p, L)$ is inherently discrete. For small $L$, the spacing between the allowed values is large. As $L$ increases, these spacings decrease, and the discrete distribution approaches a continuous envelope. To formalize this convergence, let $\tilde{\mathbb{P}}(\Delta E; p, L)$ denote the continuous distribution that one would obtain in the thermodynamic limit. We define the maximum pointwise discrepancy as
\begin{equation}
    \epsilon(L)=\sup_{\Delta E}\left|\mathbb{P}(\Delta E; p, L)- \tilde{\mathbb{P}}(\Delta E; p, L) \right|.
\end{equation}

Since the spacing between quantized values scales as $L^{-\theta}$ (see Eq.~\eqref{eq:spacing}), we expect that $\epsilon(L)\sim L^{-\theta}$ as $L\rightarrow\infty$. In practice, however, the discrete supports for different $L$ may not align pointwise even after rescaling. To address this issue, one may consider comparisons over local neighborhoods rather than exact pointwise matches. The Wasserstein metric, defined in terms of the CDFs, provides a robust measure of the “cost” required to rearrange the probability mass of one distribution into that of another \cite{Villani2009}. Formally, for two probability measures $\alpha$ and $\beta$, the first Wasserstein distance is defined as
\begin{equation}
    W_{1}(\alpha,\beta)=\inf_{\gamma\in\Gamma(\alpha,\beta)}\int_{\mathbb{R}^{2}}|u-v|d\gamma(u,v),
\end{equation}
where $\Gamma(\alpha,\beta)$ is the set of all joint distributions with marginals $\alpha$ and $\beta$. In our context, we compute the rescaled distributions $\hat{P}_{L}(u)=L^{\theta}\mathbb{P}(L^{\theta}u;p_{c},L)$, and we provide numerical evidence that $\hat{P}_{L}(u)$ converges to the universal function $G(u)$ using the Wasserstein metric.

\subsection{Convergence of rescaled probability distribution to envelope}\label{app:convergence_envelope}

\begin{figure}[h]
    \centering
    \includegraphics[width=\linewidth]{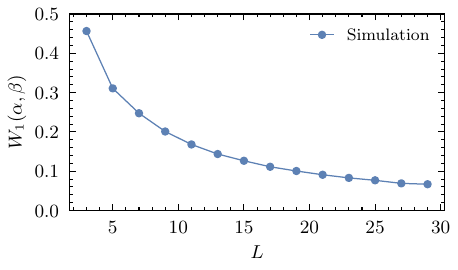}
    \caption{Plot of the Wasserstein distance between the rescaled probability distributions for adjacent odd lattice sizes $L$ and $L+2$ at criticality ($p=p_{c}$). We denote ${\alpha=\mathbb{P}(\Delta E(p_{c},L)/\Delta E_{0}(p_{c},L))}$, and ${\beta=\mathbb{P}(\Delta E(L+2,p_{c})/\Delta E_{0}(L+2,p_{c}))}$. The decreasing trend of the Wasserstein metric with increasing $L$ provides evidence that the rescaled distributions converge toward a continuous universal envelope as $L\rightarrow\infty$.}
    \label{fig:Wasserstein}
\end{figure}

Having established the expected scaling form for $\mathbb{P}_{\mathrm{fail}}(\Delta E; p, L)$, we next test whether the rescaled distributions at $p=p_{c}$ converge to a single universal envelope as $L\to\infty$. We assess the convergence of the rescaled distributions of ground state energy costs by computing the Wasserstein distance between adjacent odd lattice sizes $L$ and $L+2$. Fig.~\ref{fig:Wasserstein} shows that the Wasserstein distance decreases with increasing $L$, providing evidence that the rescaled distributions converge to a continuous limit as $L\rightarrow\infty$.

\subsection{Asymptotic behaviour of the scaling function $G(x)$}\label{app:asymptotics_gx}

Having verified that the rescaled energy-cost distributions converge to a universal form at criticality, we now consider the constraints this scaling imposes on the functional form of $G(x)$ in Eq.~\eqref{eq:FSS_general_erf_ansatz}, since these constraints restrict the class of admissible phenomenological models. In particular, the limits $x\to\pm\infty$ correspond to the deep ordered and disordered regimes, and the behaviour of $G(x)$ in these regimes determines whether the phenomenological error-function model satisfies the physically required limits of $\mathbb{P}_{\mathrm{fail}}$.

For Eq.~\eqref{eq:FSS_general_erf_ansatz} to meet the required limits, the asymptotic behaviour of $G(x)$ is constrained as follows:
\begin{itemize}
    \item For $x \to -\infty$:
    \begin{align*}
        f(x) \to 0 \quad &\implies \quad \mathrm{erf}\left(\frac{G(x)}{\sqrt{2}}\right) \to 1 \\
        &\implies \quad G(x) \to +\infty.
    \end{align*}
    \item For $x \to +\infty$:
    \begin{align*}
        f(x) \to 1/2 \quad &\implies \quad \mathrm{erf}\left(\frac{G(x)}{\sqrt{2}}\right) \to 0 \\
        &\implies \quad G(x) \to 0.
    \end{align*}
\end{itemize}
Thus, $G(x)$ should diverge positively for large negative $x$ and approach zero for large positive $x$. Standard polynomial forms (e.g., finite-order Taylor expansions) for $G(x)$ do not generally satisfy both these conditions.

\subsection{Comparison of alternative scaling ans\"{a}tze performance}\label{app:performance_comparison}

To assess the robustness of our conclusions, we compare the phenomenological error function model introduced in Sec.~\ref{subsec:nearthresh_nopostselect} against alternative finite-size scaling ans\"{a}tze. These comparisons test whether the observed scaling behaviour and extracted parameters are sensitive to the specific functional form assumed for $\mathbb{P}_{\mathrm{fail}}(p,L)$ at both $T=0$ (using MWPMD) and at $T=T_{N}(p)$ (using optimal MLD).

While both fitting procedures yield excellent data collapse and validate the finite-size scaling hypothesis, the significant difference in the extracted parameters ($B_{i}$ vs. $A_{i}$, as well as $p_{c}$ and $\nu$) is not unexpected. The two methods are sensitive to different features of the underlying energy cost distributions $\mathbb{P}(\Delta \mathcal{C};p, L)$. The first method, which directly fits the ratio $\mu/\sigma$, relies on the first two moments of the entire distribution. Its parameters are thus influenced by the global shape of $\mathbb{P}(\Delta \mathcal{C})$, including its central bulk and positive tail. In contrast, the second method fits the logical failure rate, $\mathbb{P}_{\mathrm{fail}}$, which is determined solely by the integral over the negative tail of the distribution ($\Delta \mathcal{C}<0$). This discrepancy suggests that our model of a perfect Gaussian distribution is a powerful but imperfect approximation. Any deviation from a true Gaussian, such as skewness or non-Gaussian tails, will affect the globally-calculated $\mu$ and $\sigma$ differently than it affects the area of the negative tail. Consequently, the phenomenological fit to $\mathbb{P}_{\mathrm{fail}}$ provides a more direct and reliable model for the quantity of interest (the logical error rate) even if the parameters differ from those derived from the moments of the full, approximate distribution.

To provide a more comprehensive quantitative assessment of our phenomenological error function framework, we compare its performance (using both linear and quadratic $G(x)$) against two alternative functional forms. The first alternative is a second-order polynomial in the scaling variable $x=(p-p_c)L^{1/\nu}$, augmented with a leading-order finite-size correction term dependent on $L$:
\begin{align}\label{eq:poly_L_ansatz_revised} 
    f_{\mathrm{poly-L}}(p, L) &= A_{\mathrm{poly-L}}x^2 + B_{\mathrm{poly-L}}x\\ &+ C_{\mathrm{poly-L}} + D_{\mathrm{poly-L}}L^{-1/\mu},
\end{align}
where $A_{\mathrm{poly-L}}, B_{\mathrm{poly-L}}, C_{\mathrm{poly-L}}, D_{\mathrm{poly-L}}$, along with $p_{c}$, $\nu$ (which define $x$), and $\mu$ are fitting parameters, totaling 7 free parameters \cite{WangChenyang2003Ctia,Watson_2014}. This model attempts to capture the scaling behaviour and leading corrections without imposing the error function structure.

The second alternative is a second-order polynomial without the $L$-dependent correction.
\begin{equation}\label{eq:exp_FSS_reiterated_revised}
    f_{\mathrm{poly}}(p, L) = A_{\mathrm{poly}}x^2 + B_{\mathrm{poly}}x + C_{\mathrm{poly}},
\end{equation}
where $A_{\mathrm{poly}}, B_{\mathrm{poly}},  C_{\mathrm{poly}}$, along with $p_c$ and $\nu$ are fitting parameters, totaling 5 free parameters.

We compare these ansatz using the Residual Sum of Squares (RSS), Akaike Information Criterion (AIC), and Bayesian Information Criterion (BIC) as metrics. Lower values for these metrics generally indicate a better fit, with AIC and BIC penalizing models with more parameters. We compare all four ans\"{a}tze by fitting them to the full dataset of logical failure rates $\mathbb{P}_{\mathrm{fail}}(p,L)$ at $T=0$ and $T=T_{N}(p)$ for odd $L$ between $L_{\mathrm{min}}$ and $L_{\mathrm{max}}$ ($L=9$ to $31$) and $p$ values spanning the critical region ($p=0.08$ to $0.12$ for $T=0$ and $p=0.09$ to $0.13$ for $T=T_{N}(p)$). The results are summarized in Tabs.~\ref{tab:ansatz_compare_all_p_revised} and \ref{tab:ansatz_compare_all_p_revised_optimal}.

\begin{table*}[t]
    \begin{ruledtabular}
    \begin{tabular}{lccccc}
    Model Ansatz & RSS & AIC & BIC & $k$ & $N$ \\
    \hline
    Quadratic Erf Model & $2.313\times10^{-4}$ & -2782.71 & -2766.12 & 5 & 204 \\
    Linear Erf Model & $1.364\times10^{-2}$ & -1953.01 & -1939.73  & 4 & 204 \\
    Polynomial-L (Eq.~\eqref{eq:poly_L_ansatz_revised}) & $3.132\times10^{-3}$   & -2247.19  & -2223.96   & 7 & 204 \\
    Simple polynomial (Eq.~\eqref{eq:exp_FSS_reiterated_revised}) & $3.182\times10^{-3}$ &  -2247.95     & -2231.36     & 5 & 204 \\
    \end{tabular}
    \end{ruledtabular}
    \caption{\label{tab:ansatz_compare_all_p_revised}Comparison of fit-quality metrics for the four ans\"{a}tze fitted to the full dataset ($L \in [9, 31]$, all $p$ near $p_c$ with $T=0$), where $k$ is the number of free parameters, and $N$ is the number of data points.}
\end{table*}

\begin{table*}[t]
    \begin{ruledtabular}
    \begin{tabular}{lccccc}
    Model Ansatz & RSS & AIC & BIC & $k$ & $N$ \\
    \hline
    Quadratic Erf Model & $3.218\times10^{-4}$ & -2715.39 & -2698.80 & 5 & 204 \\
    Linear Erf Model & $1.524\times10^{-2}$ & -1930.33 & -1917.06  & 4 & 204 \\
    Polynomial-L (Eq.~\eqref{eq:poly_L_ansatz_revised}) & $2.749\times10^{-3}$   & -2273.80  & -2250.58   & 7 & 204 \\
    Simple polynomial (Eq.~\eqref{eq:exp_FSS_reiterated_revised}) & $2.748\times10^{-3}$ &  -2277.84     & -2261.25     & 5 & 204 \\
    \end{tabular}
    \end{ruledtabular}
    \caption{\label{tab:ansatz_compare_all_p_revised_optimal}Comparison of fit-quality metrics for the four ans\"{a}tze fitted to the full dataset ($L \in [9, 31]$, all $p$ near $p_c$ with $T=T_{N}(p)$), where $k$ is the number of free parameters, and $N$ is the number of data points.}
\end{table*}

\section{Assumptions and scope of applicability}\label{app:scope}

The analyses in the main text are intended to apply to two-dimensional topological stabilizer codes with local checks and growing distance. For such codes, nontrivial logical operators can be represented by connected string-like operators \cite{10.1063/5.0021068,Bombin_2012}. Under the standard mapping to a classical statistical mechanical model, these logical strings correspond to system-spanning domain walls. This is the central structural assumption underlying our four-regime picture: logical failure is controlled by the free energy cost of inserting such a domain wall.

We assume that the physical noise is described by an i.i.d. Pauli channel, or more generally by local Pauli error probabilities satisfying the Nishimori conditions. Under this assumption, maximum-likelihood decoding is equivalent to comparing free energies of the corresponding classical statistical mechanical model along the Nishimori line. In particular, the posterior weight of each logical sector is proportional to the partition function in the corresponding domain wall sector.

Below threshold, we assume that the associated statistical mechanical model is in an ordered regime. In this regime, inserting a nontrivial logical domain wall has a free energy cost whose typical scale grows macroscopically with the system size. For the two-dimensional codes considered here, the expected scaling is linear,
\[
    \Delta F(\mathcal{L}_{m}) \sim \sigma_m(p)L + o(L),
\]
with $\sigma_m(p)>0$ for each independent logical operator $\mathcal{L}_{m}$. Logical failure is controlled by the lower tail of this domain wall free energy distribution, corresponding to disorder and thermal fluctuations that make a nontrivial logical sector more favourable than the trivial sector. Thus, below threshold, reliable storage follows from a macroscopic domain wall cost together with sufficiently suppressed lower-tail fluctuations.

Above threshold, we assume that this macroscopic free energy separation between logical sectors disappears. Equivalently, system-spanning domain walls have at most subextensive free energy cost in the thermodynamic limit. The logical sectors then become asymptotically indistinguishable, and, under the natural symmetry assumption that their posterior weights become uniform over the $K$ logical sectors, the logical failure rate approaches the trivial value
\[
    \mathbb{P}_{\mathrm{fail}}\to \frac{K-1}{K}.
\]

For the finite-size scaling analysis near threshold, we further assume that the relevant order-disorder transition is continuous, with a diverging correlation length
\[
    \xi \sim |p-p_c|^{-\nu}.
\]
In this case, the logical failure rate is expected to obey the scaling form
\[
    \mathbb{P}_{\mathrm{fail}}(p,L) = f\!\left((p-p_c)L^{1/\nu}\right),
\]
up to finite-size corrections. The value of $\nu$ is determined by the universality class of the associated statistical mechanical model. For example, surface code variants under bit-flip noise map to the two-dimensional random-bond Ising model, while colour codes map to disordered three-body Ising models \cite{PhysRevA.77.042322,landahl2011faulttolerantquantumcomputingcolor}. Thus different code families may share the same four-regime structure while having different critical exponents and scaling functions.

The framework therefore applies most directly to two-dimensional topological stabilizer codes with perfect measurements, including surface codes, colour codes, and related homological quantum LDPC constructions. Such models often have local checks, string-like logical operators, and associated statistical mechanical models with an ordered and disordered regime which are separated by a second-order phase transition. These results may also apply to families such as bivariate bicycle codes when their fixed local check structure gives rise to surface code-like statistical mechanical behaviour \cite{chen2025anyontheorytopologicalfrustration,liang2025generalizedtoriccodestwisted}. Extensions to higher-dimensional codes, nonlocal interactions, or models with first-order transitions require separate analysis and are not assumed in the arguments of this work. Nonetheless, we observed in Section~\ref{sec:circuit-level} that the results developed earlier can be applied to models in which the order of the phase transition is unknown. We leave an investigation of whether this phase transition is indeed second order to future work.

\end{document}